\documentclass[letterpaper, 10 pt, conference]{ieeeconf}  % Comment this line out if you need a4paper

\usepackage[pdftex,dvipsnames]{xcolor}  % Coloured text etc.
\usepackage[colorinlistoftodos,prependcaption,textsize=tiny]{todonotes}

\IEEEoverridecommandlockouts                              % This command is only needed if you want to use the \thanks command
\usepackage{amsthm}
\usepackage{graphics} % for pdf, bitmapped graphics files
\usepackage{epsfig} % for postscript graphics files
\usepackage{times} % assumes new font selection scheme installed
\usepackage{amsmath} % assumes amsmath package installed
\usepackage{amssymb}  % assumes amsmath package installed

\usepackage[maxbibnames=99,giveninits=true]{biblatex}
\usepackage{bm}
\usepackage{bbm}
\usepackage{xcolor}
\usepackage{graphicx}
\usepackage{algorithm}
\usepackage[noend]{algpseudocode}
\usepackage{booktabs}
\usepackage{siunitx}% An awesome package for typesetting and manipulation numbers and units.
\usepackage{hyperref}% Used for insert the link

\usepackage{caption}% Better control over caption
\usepackage{lipsum}% Example text
\usepackage{multirow}
\usepackage{diagbox} 
\usepackage{subcaption}

\newcommand{\R}{\mathbb{R}}%commands for easy math notations

\newtheorem{definition}{\bf Definition}
\newtheorem{assumption}{\bf Assumption}
\newtheorem{theorem}{\bf Theorem}
\newtheorem{lemma}{\bf Lemma}

\begin{document}
\title{\LARGE \bf
Quantification of Distributionally Robust Risk of Cascade of Failures \\in Platoon of Vehicles
}

\author{Vivek Pandey, Guangyi Liu, Arash Amini, and Nader Motee 
\thanks{
 V. Pandey, G. Liu, A. Amini and N. Motee are with the Department of Mechanical Engineering and Mechanics, Lehigh University, Bethlehem, PA, 18015, USA. {\tt\small \{vkp219, gliu, ara416, motee\}@lehigh.edu}.\endgraf
}
}

\maketitle

% Replace this two for submission
% \thispagestyle{plain}
% \pagestyle{plain}

\thispagestyle{empty}
\pagestyle{empty}

\begin{abstract} 
Achieving safety is a critical aspect of attaining autonomy in a platoon of autonomous vehicles. In this paper, we propose a distributionally robust risk framework to investigate cascading failures in platoons. To examine the impact of network connectivity and system dynamics on the emergence of cascading failures, we consider a time-delayed network model of the platoon of vehicles as a benchmark. To study the cascading effects among pairs of vehicles in the platoon, we use the measure of conditional distributionally robust functional. We extend the risk framework to quantify cascading failures by utilizing a bi-variate normal distribution. Our work establishes closed-form risk formulas that illustrate the effects of time-delay, noise statistics, underlying communication graph, and sets of soft failures. The insights gained from our research can be applied to design safe platoons that are robust to the risk of cascading failures. We validate our results through extensive simulations.
\end{abstract}

%%%%%%%%%%%%%%%%%%%%%%%%%%%%%%%%%%%%%%%%%%%%%%%%%%%%%%%%%%%%%%%%%%%%%%%%%%%%%%%%%%%%%

\section{Introduction}
Networked systems are omnipresent and play a vital role in various engineering applications, ranging from power networks to platoon of autonomous cars. However, the performance of network dynamical systems is often affected by communication issues and external disturbances. These disturbances can steer the system from an optimal state of operation towards suboptimal conditions or even complete network failure. This phenomenon is frequently observed in power networks, supply chains, and financial systems, where events leading to system breakdown occur frequently \cite{acemoglu2015systemic, bamieh2012coherence, bertsimas1998air, dorfler2012synchronization,kessler1989analysis}.\\
The intricacy of network systems often comes with fragility, which can result in a complete breakdown of the system. In consensus networks, where a group of agents negotiates and decides on a particular action, the existence of time delays and environmental noise can cause an agent to deviate from the consensus. The presence of uncertainty in these networks has piqued researchers' interest in the risk analysis of complex dynamical systems \cite{zhang2018cascading, zhang2019robustness}. Investigating these phenomena is crucial from a system design standpoint, as the failure of one agent may trigger a domino effect throughout the network. 

In this paper, we develop a theoretical framework to quantify distributionally robust risk to evaluate the effect of cascading  failures in the consensus network. Our objective is to develop a risk framework robust to uncertainty in the probability measures characterizing the stochastic disturbance.  We also aim to highlight how the failures in one part of the network affects the safety of other components in the network. Quantification of such phenomena offers valuable insight in designing a reliable and robust network that can minimize the risk of failure of the system or allow the system to take preventive measures to nullify such effects. \\
{\it Our Contributions:} Continuing with our previous work on the risk analysis of networked systems \cite{somarakis2018risk, ,liu2021risk,liu2022risk}, this paper is a first step towards developing a distributionally robust cascading risk framework. 
 % First, we define risk using conditional distributionally robust  functional using steady state statistics of time - delayed vehicle platoon system.
 First, we employ conditional distributionally robust  functional to define risk which is later quantified using steady state statistics of time-delayed system of platoon of vehicles. In particular, explicit formula is obtained for distributionally robust cascading risk. 
% given that other vehicles lie in a certain set of inter-vehicle distance. 
We also provide a lower bound for the distributionally robust cascading risk as function of eigenvalues of covariance matrix. Finally, with extensive simulations, we present the cascading risk profile of vehicles in platoon for different scenarios.

%%%%%%%%%%%%%%%%%%%%%%%%%%%%%%%%%%%%%%%%%%%%%%%%%%%%%%%%%%%%%%%%%%%%%%%%%%%%%%%%%%%%
\section{Mathematical Notation}
We denote the non-negative orthant of the Euclidean space $\R^n$ by $\mathbb{R}_{+}^n$, its set of standard Euclidean basis by $\{\bm{e}_1, \dots, \bm{e}_n\}$,  and  the vector of all ones by $\bm{1}_n = [1, \dots, 1]^T$. The indicator function of a set $A$ is represented by $\bm{1}_{[A]}$. We utilize the notation $\Tilde{\bm{e}}_i = \bm{e}_{i + 1} - \bm{e}_i$ for all $i \in \{1, \cdots, n - 1\}.$ We reserve the notation $S^n_{+}$ , $S^n_{++}$ to denote the cone of symmetric positive semidefinite and symmetric positive definite $n \times n$ matrices respectively. For every $X_1, X_2 \in S^n_{++} $, we write $X_2 \preceq X_1$ if and only if $ X_1 - X_2 \in S^n_{+}$. The $n \times n$ identity matrix is denoted by $I_n$

\vspace{0.1cm}
{\it Algebraic Graph Theory:} A weighted graph is defined by $\mathcal{G} = (\mathcal{V}, \mathcal{E}, \omega)$, where $\mathcal{V}$ is the set of nodes, $\mathcal{E}$ is the set of edges (feedback links), and $\omega: \mathcal{V} \times \mathcal{V} \rightarrow \mathbb{R}_{+}$ is the weight function that assigns a non-negative number (feedback gain) to every link. Two nodes are directly connected if and only if $(i,j) \in \mathcal{E}$.

\begin{assumption}  \label{asp:connected}
    Every graph in this paper is connected. In addition, for every $i,j \in \mathcal{V}$, the following properties hold:
    \begin{itemize}
        \item $\omega(i,j) > 0$ if and only if $(i,j) \in \mathcal{E}$.
        \item $\omega(i,j) = \omega(j,i)$, i.e., links are undirected.
        \item $\omega(i,i) = 0$, i.e., links are simple.
    \end{itemize}
    
\end{assumption}

The Laplacian matrix of $\mathcal{G}$ is a $n \times n$ matrix $L = [l_{ij}]$ with elements
\[
    l_{ij}:=\begin{cases}
        \; -k_{i,j}  &\text{if } \; i \neq j \\
        \; k_{i,1} + \ldots + k_{i,n}  &\text{if } \; i = j 
    \end{cases},
\]
where $k_{i,j} := \omega(i,j)$. Laplacian matrix of a graph is symmetric and positive semi-definite \cite{van2010graph}. Assumption \ref{asp:connected} implies the smallest Laplacian eigenvalue is zero with algebraic multiplicity one. The spectrum of $L$ can be ordered as 
$
    0 = \lambda_1 < \lambda_2 \leq \dots \leq \lambda_n.
$
The eigenvector of $L$ corresponding to $\lambda_k$ is denoted by $\bm{q}_{k}$. By letting $Q = [\bm{q}_{1} | \dots | \bm{q}_{n}]$, it follows that $L = Q \Lambda Q^T$ with $\Lambda = \text{diag}[0, \lambda_2, \dots, \lambda_n]$. We normalize the Laplacian eigenvectors such that $Q$ becomes an orthogonal matrix, i.e., $Q^T Q = Q Q^T = I_{n}$ with $\bm{q}_1 = \frac{1}{\sqrt{n}} \bm{1}_n$. 

\vspace{0.1cm}
{\it Probability Theory:} Let $\mathcal{L}^{2}(\mathbb{R}^{q})$ be the set of all $\R^q-$valued random vectors $\bm{z} = [z^{(1)}, \dots ,z^{(q)}]^T$ of a probability space $(\Omega, \mathcal{F}, \mathbb{P})$ with finite second moments. A normal random variable $\bm{y} \in \mathbb{R}^{q}$ with mean $\bm{\mu} \in \mathbb{R}^{q}$ and $q \times q$ covariance matrix $\Sigma$ is represented by $\bm{y} \sim \mathcal{N}(\bm{\mu}, \Sigma)$. The error function erf: $\mathbb{\R} \rightarrow (-1,1)$ is erf$(x) = \dfrac{2}{\sqrt{\pi}} \int_{0}^{x} e^{-t^2} dt.$ We employ standard notation $\text{d} \bm{\xi}_t$ for the formulation of stochastic differential equations.

%%%%%%%%%%%%%%%%%%%%%%%%%%%%%%%%%%%%%%%%%%%%%%%%%%%%%%%%%%%%%%%%%%%%%%%%%%%%%%%%%%%%%
\section{Problem Statement}\label{problemstatement}

\begin{figure}[t]
    \centering
	\includegraphics[width=\linewidth]{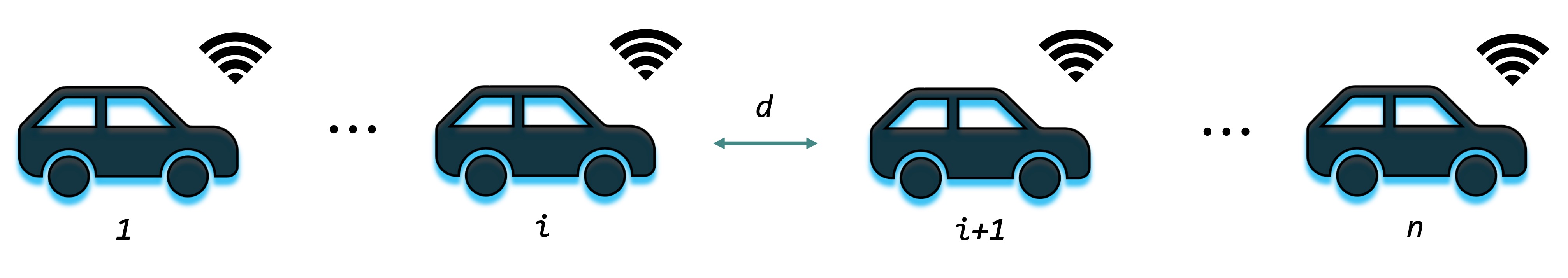}
	\caption{Schematic of a platoon. Vehicles try to maintain a fixed distance $d$ and achieve the same velocity by updating the feedback using the communication network.}
	\label{fig:platoon_schematics}
\end{figure}
% We consider a system of platoon of $n$ vehicles moving along straight line such that the dynamics of the $ith$ vehicle in the platoon is given by the following second order stochastic differential equations,

We consider a system of platoon of $n$ vehicles moving along straight line as shown in figure \ref{fig:platoon_schematics}. The vehicles are labeled in descending order such that the $n$'th vehicle is the leader of the platoon. We denote the state of the $i$'th vehicle by $[x_t^{(i)};v_t^{(i)}]$, where $x_t^{(i)}$ and $v_t^{(i)}$  are the position and velocity of the $i$'th vehicle respectively. The second order dynamics of $i$'th vehicle in the platoon is expressed by the following first order linear stochastic differential equations,

\begin{equation} \label{eq:dyn_vehicle}
\begin{aligned}
    d{x}_t^{(i)} &= {v}_t^{(i)} dt \\
    d{v}_t^{(i)} &={u}_t^{(i)} dt  + g d{\xi}_t^{(i)}, 
\end{aligned}
\end{equation}
where ${u}_t^{(i)}$ is the control input of the $i$'th vehicle. We model the exogenous disturbance from the environment entering the dynamics of the $i$'th vehicle using the additive term $g d{\xi}_t^{(i)}$, where $d{\xi}_t^{(i)}$ is the differential Brownian motion\footnote{The stochastic process ${\xi}_t^{(i)} \in \mathcal{L}^2(\R)$ denotes the real valued Weiner process.} and $g \in \R_{++}$ is the diffusion coefficient independent of the state and control input of the $i$'th vehicle. The diffusion coefficient $g$ which, is a measure of the covariance of the noise entering the  vehicles is assumed identical for all the vehicles. We define the vehicles to form a platoon if they satisfy the following conditions in the steady state: (i) the pairwise difference between the position coordinate of every two consecutive vehicles
% $|({x}_t^{(i+1)} - {x}_t^{(i)}) - ({x}_t^{(j+1)} - {x}_t^{(j)})|$ 
converges to zero for all $i,j \in \{1, \cdots, n\}$; and (ii) all vehicles attain the same constant velocity in the steady state. 
The control input that satisfies the previous two conditions is determined the following feedback control law \cite{yu2010some}
\begin{multline}     \label{eq:control_law}
       {u}_t^{(i)} = \sum_{j = 1}^{n}k_{ij}\big(v^{j}_{t - \tau} - v^{i}_{t - \tau}\big)
    \\+ \beta \sum_{j = 1}^{n}k_{ij}\big(x^{j}_{t - \tau} - x^{i}_{t - \tau} -(d_j - d_i)\big).
\end{multline}
The parameter $\beta$ is a weighting parameter that assigns relative importance to the velocity and position term in the control law. In order to account for the non - zero time in the communication channel between the vehicles, we assume that the states in control input are delayed. We denote this time delay using the parameter $\tau \in \R_{++}$ which is assumed constant for all the vehicles. 

In order to write the dynamics of the platoon, we introduce the following notations. We define a vector of positions, velocities and exogenous noise input as $\bm{\text{x}}_t = [{x}_t^{(1)}, \cdots, {x}_t^{(n)}]^T$, $\bm{\text{v}}_t = [{v}_t^{(1)}, \cdots, {v}_t^{(n)}]^T$, and $\bm{\xi_t} = [{\xi}_t^{(1)}, \cdots, {\xi}_t^{(n)}]^T$ respectively. We define the vector of target distance from the leader of the platoon $\bm{\text{y}} = [d, 2d, \cdots, nd]^T $, where $d \in  \R_{++}$. Applying the control law \eqref{eq:control_law}
in \eqref{eq:dyn_vehicle} for all $i = 1 \cdots n$, the closed loop dynamics of the network of $n$ vehicles with graph Laplacian $L$, can be represented by the following $2n$ first order stochastic delayed differential equations,
\begin{equation} \label{eq:dyn_platoon}
\begin{aligned}
    d\bm{\text{x}}_t &= \bm{\text{v}}_t dt \\
    d\bm{\text{v}}_t &=  -L \bm{\text{v}}_{t-\tau} dt - \beta L (\bm{\text{x}}_{t-\tau} - \bm{\text{y}}) dt + g d\bm{{\xi}_t}, 
 \end{aligned}
\end{equation}
for all $t \geq 0$ and given deterministic initial conditions for $\bm{\text{x}}_t$ and $\bm{\text{v}}_t$ for $t \in [-\tau, 0]$. The standard result \cite{mohammed1984stochastic}, \cite{Evans2013} guarantees  that the states $\{ (\bm{\text{x}}_{t}, \bm{\text{v}}_{t})\}_{t\geq -\tau}$ of  stochastic delayed differential equation \eqref{eq:dyn_platoon} represent a well defined stochastic process.\\

The problem is to quantify the distributionally robust risk of a cascade of failures as a function of communication graph Laplacian, time-delay and statistics of noise, conditioned on the fact that at another location, the inter-vehicle distance of consecutive pair of vehicles lies within a certain set. 
% We refer to the inter-vehicle lying within a certain set as the pair having undergone soft failure.
Since the exact knowledge of probability measures characterizing the stochastic disturbance is not known in practice,  it becomes important to develop risk models robust to changes in noise statistics. 
To this end, we develop a distributionally robust systemic risk framework to study the cascade of failures using the steady state statistics of the closed loop system \eqref{eq:dyn_platoon}. The rest of the paper is organized as follows. In \S \ref{prelims}, we state the stability conditions for the platoon, obtain steady state statistics required to quantify risk, and present a formal framework of distributionally robust risk measure. These results help us to quantify the distributionally robust cascading risk for the platoon of vehicles in \S \ref{sec:dist_robust_risk}, which constitutes the main contribution of this work. In \S \ref{sec:case_studies}, we examine risk profile  simulation results for some special graph topologies. The proofs of all theoretical results are provided in the Appendix.

%%%%%%%%%%%%%%%%%%%%%%%%%%%%%%%%%%%%%%%%%%%%%%%%%%%%%%%%%%%%%%%%%%%%%%%%%%%%%%%%%%%%%
\section{Preliminary Results}\label{prelims}

In this section, we present the conditions for the stability of platoon, and the steady state statistics of the inter-vehicle distance which is later utilized in the risk analysis. We define a formal framework of distributionally robust cascading risk measure using conditional distributionally robust functionals.
% and an bound on the conditional distributionally robust functional

\subsection{Stability of the Deterministic Platoon}
We define the convergence of the platoon if the steady-state distance between pair of consecutive vehicles attain the same constant value $d$, and the velocity of all the vehicles reach the same constant value. 
These conditions can be equivalently stated as $\lim_{t \rightarrow \infty}|v_t^{(i)} - v_t^{(j)}| = 0$ and $\lim_{t \rightarrow \infty}|x_t^{(i)} - x_t^{(j)} - (i - j)d| = 0$. 
Using the result in \cite{yu2010some} \cite{bellman1963_dde}, we conclude that the deterministic platoon converges if and only if 
\begin{equation}\label{eq:platoon_stability}
    (\lambda_i\tau, \beta\tau) \in S,
\end{equation}
where $\lambda_i, i = 1, \cdots, n$ are the eigenvalues of the graph Laplacian, and the set S is as defined below, 
\begin{multline*}
    S = \bigg\{(s_1, s_2) \in \R^2 | s_1 \in \big(0, \frac{\pi}{2}\big), s_2 \in \bigg(0, \frac{a}{\tan(a)},
    \\\text{for} ~ a  \in \big(0, \frac{\pi}{2}\big)\bigg), \text{the solutions of a}\sin(a) = s_1 \bigg\}.
\end{multline*}

The results in this paper are derived when the closed loop dynamics of the platoon is stable, i.e., when \eqref{eq:platoon_stability} holds. 

\subsection{Steady State Statistics of inter-vehicle Distances}
Since the graph Laplacian $L \in S^n_+$, we utilize the spectral decomposition $L = Q \Lambda Q^T$ to decouple \eqref{eq:dyn_platoon} using the following coordinate transformation
\begin{equation}\label{eq:trans_pushforward}
\bm{z}_t = Q^T(\bm{\text{x}}_t - \bm{\text{y}}), ~\text{and} ~ \bm{v}_t = Q^T\bm{\text{v}}_t. 
\end{equation}

The closed loop dynamics of the platoon in the transformed coordinate system can be written as 
\begin{equation} \label{eq:dyn_platoon_decoupled}
\begin{aligned}
    d\bm{z}_t &= \bm{v}_t dt \\
    d\bm{v}_t &=  - \Lambda \bm{v}_{t-\tau} dt - \beta \Lambda (\bm{z}_{t-\tau} - \bm{v}) dt + g Q^T d\bm{{\xi}_t}, 
\end{aligned}
\end{equation}

The solution of \eqref{eq:dyn_platoon_decoupled} satisfying the stability condition given by \eqref{eq:platoon_stability} can be written as 

\begin{equation} \label{eq:dyn_platoon_decoupled_solution}
    \begin{bmatrix}
    {z}_t^{(i)} \\ 
    {v}_t^{(i)}
    \end{bmatrix} = \Xi({z}_{[-\tau,0]}^{(i)}; {v}_{[-\tau,0]}^{(i)}; \Phi_i(t)) + g \int_{0}^{t} \Phi_i(t-s)B_id\bm{{\xi}}_s
\end{equation}
% where $B_i = [0_{1\times n}, q_i]^T$,  $\Phi_i(t)$ is the principal solution of the deterministic platoon without time - delay, and 

where $B_i = [0_{1\times n}, q_i]^T$,  $\Phi_i(t)$ is the principal solution of the deterministic platoon, and 
% $\Xi(\cdot; \cdot; \cdot)$ is the solution of the \eqref{eq:dyn_platoon_decoupled} without any exogenous noise input \cite{Somarakis2020b}.
both $\Phi_i(t)$ and $\Gamma(\cdot; \cdot; \cdot)$ decay exponentially for all $i \in \{2, \cdots, n\}$. The steady state statistics of $\bm{z}_t$ follows a multivariate normal distribution \cite{Somarakis2020b}, i.e.,
$$
\Bar{\bm{z}} \sim \mathcal{N}(0\cdot \bm{1}_n, \Sigma_z),
$$
where
% $\bm{\mu_z} = 0\cdot \bm{1}_n $, and 
$\Sigma_z = \text{diag}\{\sigma_{z_1}^2, \cdots, \sigma_{z_n}^2\}$, in which $\sigma_{z_i}^2 = \frac{g^2 \tau^3}{2\pi}f(\lambda_i\tau,\beta\tau)$ with
\begin{equation}\label{eq:f(S_1,s_2)}
    f(s_1,s_2) = \int_{\R}^{}\frac{dr}{(s_1 s_2 - r^2 \cos(r))^2 + r^2(s_1 - r\sin(r))^2}.
\end{equation}
Inverting the transformation \eqref{eq:trans_pushforward}, we have 
\begin{equation} \label{eq:trans_pullback}
    \bm{x}_t = Q\bm{z}_t + \bm{y}.
\end{equation}
We define the steady state distance vector of the platoon as $\Bar{\bm{d}} \in \R^{n-1}$ \cite{liu2021risk} such that  

\begin{equation} \label{eq: steady_state_dist}
    \Bar{\bm{d}} = D^T Q\Bar{\bm{z}}+ d\bm{1}_{n-1},
\end{equation}
where $D = [\Tilde{\bm{e}}_1| \cdots | \Tilde{\bm{e}}_{n-1}]$. The $i$'th element in $\Bar{\bm{d}}$, i.e.,$\Bar{\bm{d}}_i$ denotes the steady state distance between the $i$'th and $(i+1)$'th vehicle in the platoon for all $i \in \{1, \cdots, n -1\}.$
Using the result in \cite{liu2021risk}, the steady state distance vector $\Bar{\bm{d}}$ follows a multivariate normal distribution 
% such that 
$$
\Bar{\bm{d}} \sim \mathcal{N}(d\bm{1}_{n-1}, \Sigma),
$$
 % = [\sigma_{ij}] 
% where
% such that $\Sigma\in \R^{(n-1) \times (n-1)}$ is given by
such that $\Sigma\in S^{n-1}_{++}$ is given by
\begin{equation}\label{eq:sigma_transform}
    \Sigma = D^TQ\,\mathbb{E}[\Bar{\bm{z}}\, \Bar{\bm{z}}^T]\,Q^TD,
\end{equation}
% where $F = \text{diag}\{f(\lambda_1\tau,\beta \tau), \cdots, f(\lambda_n\tau,\beta \tau)\}.$
In terms of individual elements, $\Sigma = [\sigma_{ij}]$ is given as 
% where $\Sigma = [\sigma_{ij}] \in \R^{(n-1) \times (n-1)}$ is given elementwise as
\begin{equation}\label{eq:steady_state_cov}
    \sigma_{ij} = g^2 \frac{\tau^3}{2\pi}\sum_{k = 1}^{n}(\Tilde{\bm{e}}_i^Tq_k)(\Tilde{\bm{e}}_j^Tq_k)f(\lambda_k\tau,\beta \tau),
\end{equation}
for all $i,j \in \{1, \cdots, n-1\}$ and $(\lambda_k\tau,\beta \tau)$ as defined in \eqref{eq:f(S_1,s_2)}. For simplicity of notation, we denote the diagonal elements $\sigma_{ii}$ as $\sigma_{i}^2$.

\subsection{Systemic Event and Systemic Set}
A {\it systemic} event is a failure that will potentially lead to an overall malfunction of the network \cite{Somarakis2020b,liu2021risk}. In probability space $(\Omega, \mathcal{F}, \mathbb{P})$, the set of systemic events of random variable $y: \Omega \rightarrow \R$ is defined as $\{ \omega \in \Omega ~|~y(\omega) \in \mathcal{W}^*\}$. We define a collections of supersets $\{\mathcal{W}_{\delta}~|~\delta \in [0,\infty]\}$ of $\mathcal{W}^*$ that satisfy the following conditions for any sequence $\{\delta_n\}^{\infty}_{n=1}$ with property $\lim_{n \rightarrow \infty} \delta_n = \infty$ 
\begin{itemize}
    \item $\mathcal{W}_{\delta_2} \subset \mathcal{W}_{\delta_1}$ when $\delta_1 < \delta_2$
    \item $\lim_{n \rightarrow \infty} \mathcal{W}_{\delta_n} = \bigcap_{n=1}^{\infty} \mathcal{W}_{\delta_n} = \mathcal{W}^*$.
\end{itemize}
Suppose that the knowledge of a soft failure is given priory such that the severity of the corresponding event can be represented by $\delta = \inf \{\delta ~| ~ y \in \mathcal{W}_{\delta}\}$, the value of $\delta$ implies how dangerously the current state $y$ is close to the systemic event. Using the systemic level sets to represent the existing soft failures allows for a more systematic and rigorous risk assessment and management approach. By defining different risk levels, one can evaluate the trade-offs between system performance and safety and choose an appropriate level of risk based on the specific application and requirements.

We denote the sigma-algebra generated by a particular superset $\mathcal{W}_{\delta_i}$ as $\mathcal{G}_i \subseteq \mathcal{F}$ such that $\mathcal{G}_i = \{\phi, \mathcal{W}_{\delta_i}, \mathcal{W}_{\delta_i}^c, \Omega\}$, where $\mathcal{W}_{\delta_i}^c$ is the complement of the set $\mathcal{W}_{\delta_i}$.

\subsection{Distributionally Robust Risk Measure}
To develop risk measure robust to uncertainty in probability distribution of a random variable, we quantify a set of probability distributions using the notion of ambiguity set to as below \ref{def: ambig_set}. 
% Ambiguity set 
\begin{definition}\label{def: ambig_set}
Ambiguity Set of Probability Measures \cite{shapiro2022}: Let $(\Omega, \mathcal{F})$ be measurable space. We define ambiguity set $\mathcal{M}$ as a nonempty set of  probability measures (distributions) on $(\Omega, \mathcal{F})$ contained in a ball of a specified radius centered at a reference probability measure. 
\end{definition}

% Henceforth, we refer to ambiguity set of probability measures by ambiguity set. In this work, we use two different ambiguity sets, one of which is a subset of another. We clearly state their properties before their usage.    
% Consider the probability space $(\R, \mathcal{B}_{\mathbb{\R}})$, where $\mathcal{B}_{\mathbb{\R}}$ is the Borel $\sigma-$ algebra. In this work,
 
% we consider ambiguity set of multivariate normal distribution with mean 0 and uncertain covariance matrix

For exposition of our next result, we define the input and the output ambiguity set to characterize the set of probability measure of exogenous input noise vector and steady state inter-vehicle distance vector of the system respectively. 

Consider the system
\begin{equation} \label{eq:dyn_platoon_gen_cov}
\begin{aligned}
    d\tilde{\bm{\text{x}}}_t &= \tilde{\bm{\text{v}}}_t dt \\
    d\tilde{\bm{\text{v}}}_t &=  -L \tilde{\bm{\text{v}}}_{t-\tau} dt - \beta (L \tilde{\bm{\text{x}}}_{t-\tau} - \bm{\text{y}}) dt + E d\bm{{\xi}_t}, 
\end{aligned}
\end{equation}
The system governed by \eqref{eq:dyn_platoon_gen_cov} may arise if we consider communication noise in the control law. \eqref{eq:dyn_platoon_gen_cov} reduces to \eqref{eq:dyn_platoon} for $E = gI_n$.

Since the input noise vector and steady state inter-vehicle distance vector of the systems governed by \eqref{eq:dyn_platoon} and \eqref{eq:dyn_platoon_gen_cov} are normally distributed, we consider ambiguity set of multivariate normal distribution. Furthermore, since there is no ambiguity in mean of input noise which is zero, we quantity our ambiguity set in terms of covariance matrix.

% with mean $0$ 
% \begin{Lemma}
% The steady state output of the systems \eqref{eq:dyn_platoon}\eqref{eq:dyn_platoon_gen_cov} are normal random vectors. 
% \end{Proposition}

% Unlike \eqref{eq:dyn_platoon} which has input covariance matrix, $\Gamma = g^2I$,
% \eqref{eq:dyn_platoon_gen_cov} has covariance matrix $\Gamma = EE^T$.

In the next result, we quantify ambiguity set of output $\mathcal{M}_{out}$ given the input noise ambiguity set $\mathcal{M}_{in}$. We assume the input noise has multivariate normal distribution with mean zero and uncertain covariance matrix $\Tilde{\Gamma} \in S^{n-1}_{++}$. As a result, the steady state inter-vehicle distance vector has multivariate normal distribution with mean $d\bm{1}_{n-1}$ and uncertain covariance matrix  $\Tilde{\Sigma}\in S^{n-1}_{++}$.

 \begin{lemma}\label{lem:amb_set_cone}
   For the system given by \eqref{eq:dyn_platoon_gen_cov},  with the input ambiguity set 
   \begin{equation}\label{eq: input_ambg_set}
          \mathcal{M}_{in} = \Big\{\Tilde{\Gamma}~\Big|~ (1 - \epsilon)\Tilde{\Gamma}_0 \preceq \Tilde{\Gamma} \preceq  (1 + \epsilon)\Tilde{\Gamma}_0 \Big\}
   \end{equation}
   % $\mathcal{M}_{in} = \{\Tilde{\Gamma} | I(1 - \epsilon)\Tilde{\Gamma}_0 \preceq \Tilde{\Gamma} \preceq  \Tilde{\Gamma}_0(1 + \epsilon)\}$, 
   we can quantify the output ambiguity set as  
      \begin{equation}\label{eq: output_ambg_set}
\mathcal{M}_{out} = \Big\{\Tilde{\Sigma}~\Big |~ \Tilde{\Sigma} _0(1 - \epsilon) \preceq \Tilde{\Sigma}  \preceq  \Tilde{\Sigma} _0(1 + \epsilon) \Big\},
   \end{equation}
   where $\Tilde{\Gamma}_0 \in S^{n-1}_{++}$ is the reference  covariance matrix of the input noise,
   % $\Gamma_0 = g_0^2 I$,
   $\Tilde{\Sigma}_0 \in S^{n-1}_{++}$ is the output covariance matrix corresponding  to $\Tilde{\Gamma}_0$ and $\epsilon\in (0,1)$ is a measure of radius of the ambiguity set. 
\end{lemma}

We denote the sigma-algebra generated by a particular superset $\mathcal{W}_{\delta_i}$ as $\mathcal{G}_i \subseteq \mathcal{F}$ such that $\mathcal{G}_i = \{\phi, \mathcal{W}_{\delta_i}, \mathcal{W}_{\delta_i}^c, \Omega\}$, where $\mathcal{W}_{\delta_i}^c$ is the complement of the set $\mathcal{W}_{\delta_i}$. 

In order to define our distributionally robust risk measure, we introduce the definition of conditional distributionally robust functional. 
\begin{definition}\label{def: condn_drf}
Conditional Distributionally Robust Functional\cite{shapiro2022}: Let $(\Omega, \mathcal{F})$ be measurable space and $\mathcal{M}$ be ambiguity set on $(\Omega, \mathcal{F})$. For a random variable $Z: \Omega \rightarrow \R$,  we define conditional distributionally robust functional as 
\begin{equation}\label{eq:condn_drf_defn}
\mathfrak{R}:=
    \underset{\mathbb{P} \in \mathcal{M}}{{\inf}}\mathbb{E}_{\mathbb{P}|\mathcal{G}_i}[Z] ,
\end{equation}
 where $\mathbb{E}_{\mathbb{P}|\mathcal{G}_i}[Z]$ is the conditional expectation of $Z$ given $\mathcal{G}_i$.
\end{definition}

% We define the conditionally distributionally robust functional as infimum of conditional expectation to account for the worst case scenario in our application. In the system of platoon of vehicles, the risk is maximum when the steady state inter-vehicle distance is minimum. 

We now introduce our distributionally robust risk measure for cascading failures. Motivated by \cite{Peyman2018, shapiro2022}, we define our dimensionless distributionally robust cascading risk measure as 
% inverse of conditional distributionally robust functional, 
% \begin{equation}\label{eq:drf_defn}
% \mathcal{R}^{ji} = -
%     \underset{\mathbb{P} \in \mathcal{M}}{\text{inf}}\mathbb{E}_{\mathbb{P}|\mathcal{G}}[y_j|y_{\delta_i}\in \mathcal{W}_{
%     \delta_i}].
% \end{equation}

\begin{equation}\label{eq:risk_defn}
\mathcal{R}: = \frac{\kappa}{    \underset{\mathbb{P} \in \mathcal{M}}{{\inf}}\mathbb{E}_{\mathbb{P}|\mathcal{G}_i} [Z]} + \iota,
\end{equation}
where $\kappa$ has same dimension as $Z$ and $\iota$ is a dimensionless constant. We interpret \eqref{eq:risk_defn} as risk of the random variable $Z$ given $\mathcal{G}_i$ 

% \mathbb{E}_{\mathbb{P}|\mathcal{G}}[\cdot]$ denotes the conditional expected value with respect to sigma algebra $\mathcal{G}_i$ and it can be easily interpreted as $\mathbb{E}_{\mathbb{P}}[y_j|y_i \in \mathcal{W}_{\delta_i}]:= \mathbb{E}_{\mathbb{P}}[y_j|\mathcal{G}_i] $

% The various cases of risk $\mathcal{R}$ are self explanatory. As  $\mathfrak{R}$ tends towards zero or negative, risk $\mathcal{R}$ tends towards infinity while the risk $\mathcal{R}$ decays slowly as  $\mathfrak{R}$ moves away from zero towards positive value. 

% Since, $\mathcal{R}^{ji}$ is a measure of the first moment of normal distribution, it is always bounded. $\mathcal{R}^{ji} >0$ signifies the pair $j$ has high risk of collision, while $\mathcal{R}^{ji}< 0$ implies that the vehicles are relatively safe and $\mathcal{R}^{ji} = 0$ indicates that risk profile of $j$ is not affected by $\mathcal{G}_i.$

%%%%%%%%%%%%%%%%%%%%%%%%%%%%%%%%%%%%%%%%%%%%%%%%%%%%%%%%%%%%%%%%%%%%%%%%%%%%%%%%%%%%%
\section{Distributionally Robust Cascading Risk}   \label{sec:dist_robust_risk}
In this section, we provide an explicit expression to quantify the distributionally robust cascading risk of soft failures in the platoon of vehicles for a specific type of ambiguity set.   

\subsection{Quantification of Distributionally Robust Cascading Risk}
For the platoon system \eqref{eq:dyn_platoon}, let us assume that vehicles are labelled as $1, \cdots, n$ in accordance with \ref{fig:platoon_schematics}. The main scenario assumes that the inter-vehicle distance of the $i$'th pair lies in a certain set $0 \leq \bar{d}_i \leq d$ and we study the risk for the $j$'th pair conditioned on $\bar{d}_i$. To quantify the cascading risk of soft failures, we consider that in the steady state, the inter - vehicle distance of the $i$'th pair in the platoon lies in the level set  
$$
\{\bar{d}_i \in \mathcal{W}_{\delta_i} \}  ~ \text{with} ~ \mathcal{W}_{\delta_i} = \left(-\infty, \frac{d}{\delta_i + c} \right)
$$
where $\delta_i\in \R_+$ can be interpreted as a measure of safety of the $i$'th pair of vehicles. For notational simplicity, we denote $d_i^* = \frac{d}{\delta_i + c}$, where $d_i^* \in \R_+.$ 

For the exposition of our next results, we introduce the ambiguity set that quantifies the ambiguity in the diffusion coefficient $g$. Let us assume that we have an estimate of $g$ given by $g_0$ and $g$ is specified by the set $(1 - \epsilon) g_0^2 \leq g^2 \leq (1 + \epsilon) g_0^2 $. Then using \eqref{eq:steady_state_cov}, we conclude that

% {\color{red} Please fix the typo:}
% \sigma_{\color{red} ij,0}^2

\begin{equation}\label{eq:ambiguity_set_2}
(1 - \epsilon) 
\sigma_{ij,0}^2
\leq \sigma_{ij}^2 \leq (1 + \epsilon) \sigma_{ij,0}^2, 
\end{equation}
where $\sigma_{ij,0}$ are elements of covariance matrix corresponding to $g_0$ for all $i, j \in \{1, \cdots, n-1\}.$ We denote the ambiguity set \eqref{eq:ambiguity_set_2} by $\mathcal{M}_g$. 

\begin{theorem}\label{thm: exptd_value}
Suppose that platoon \eqref{eq:dyn_platoon} reaches the steady state, and the inter-vehicle distance of the $i$'th pair, $\bar{d}_i$ belongs to the level set $\mathcal{W}_{\delta_i}$. For $i,j \in \{1, \cdots, n-1\}$ and $i \neq j$, the conditional expected value of inter-vehicle distance of $j$'th pair, $\bar{d}_j$ given $\bar{d}_i$ can be expressed as
% belongs to the level set $\mathcal{W}_{\delta_i}$ 
% $$\mathbb{E}_\mathbb{Q}[y_j | y_i \in \mathcal{W}_{\delta_i}] = \frac{|\rho|}{\sqrt{2 \pi}}\frac{\sigma_j e^{\frac{-\rho^2y^{*2}}{2 \sigma_i^2}}}{1 + erf(\frac{y^*}{\sqrt{2} \sigma_i})} $$
$$\mathbb{E}_{\mathbb{P}}[\bar{d}_j | \bar{d}_i \in \mathcal{W}_{\delta_i}] = d -\sqrt{\frac{{2}}{{\pi}}}\frac{\rho_{ji} \sigma_j e^{-\frac{(d_i^* -d)^2}{2 \sigma_i^2}}}{1 + \textup{erf}\left(\frac{d_i^* - d}{\sqrt{2} \sigma_i}\right)},
$$
where $\rho_{ji} = \frac{\sigma_{ji}}{\sigma_i \sigma_j}$ is the correlation between the  $i$'th and the $j$'th pair of vehicles.
\end{theorem}
The above theorem addresses the conditional expectation of inter-vehicle distance of $j$'th pair given that the inter-vehicle distance of the $i$'th pair is less than $d_i^* \in \R_+$. 
When there is no correlation between $i$'th and $j$'th pair of vehicles, conditional expectation is the expectation with respect to the trivial $\sigma - $algebra, since the random vector $\Bar{\bm{d}}$ is normally distributed with mean $d\bm{1}_{n-1}$, the conditional expected value equals $d$, which can also be concluded directly from the Theorem \ref{thm: exptd_value} by setting $\rho_{ji} = 0$.  

For the exposition of our next result, we introduce the notation
\begin{equation}\label{eq:h_eps}
    h(\epsilon) = d -\sqrt{\frac{2}{\pi}}\frac{\rho_{ji}\sigma_{j,0} \sqrt{ 1 + \epsilon}~ e^{\frac{-(d^* - d)^2}{2 \sigma_{i,0}^2 ( 1 + \epsilon)}}}{\left(1 + \textup{erf}\left(\frac{(d^* -d)}{\sqrt{2( 1 + \epsilon)} \sigma_{i_0} }\right)\right)}.
\end{equation}

In order to quantity the cascading risk for the system of platoon \eqref{eq:dyn_platoon}, we choose $\kappa = d$ and $\iota = -1$ such that the distributionally robust cascading risk of the $j$'th pair conditioned on $i$'th pair is 

\begin{equation}\label{eq:risk_defn_platoon}
\mathcal{R}^{ji}: = \frac{d}{\underset{\mathcal{M}_g}{{\inf}} ~ \mathbb{E}_{\mathbb{P}}[\bar{d}_j | \bar{d}_i \in \mathcal{W}_{\delta_i}]} -1.
\end{equation}

\begin{theorem}\label{thm: dr_risk}
        The conditional expected value of inter-vehicle distance derived in Theorem\ref{thm: exptd_value} is a monotonic increasing function  of $g$ for $\rho < 0$ and monotonic decreasing function of $g$ for $\rho > 0$ and for the ambiguity set \eqref{eq:ambiguity_set_2}, the distributionally robust cascading risk defined in \eqref{eq:risk_defn_platoon} is  
    % $$\underset{g}{\sup}~\mathbb{E}_\mathbb{Q} \left[y_j | y_i \in \mathcal{W}_{\delta_i} \right] = \frac{|\rho|}{\sqrt{2 \pi}}\frac{\sigma_j ( 1 + \epsilon) e^{\frac{-\rho^2y^{*2}}{2 \sigma_i^2 ( 1 + \epsilon)^2}}}{1 + erf(\frac{y^*}{\sqrt{2} \sigma_i ( 1 + \epsilon)} )} $$
% in the case of $g$ lying in $g_0\sqrt{(1 - \epsilon)}\leq g \leq g_0\sqrt{(1 + \epsilon)} $
        % $$\underset{g}{\sup}~\mathbb{E}_\mathbb{Q} \left[y_j | y_i \in \mathcal{W}_{\delta_i} \right] = -\frac{\rho}{\sqrt{2 \pi}}\frac{\sigma_j ( 1 + \epsilon) e^{\frac{-\rho^2y^{*2}}{2 \sigma_i^2 ( 1 + \epsilon)^2}}}{1 + erf(\frac{y^*}{\sqrt{2} \sigma_i ( 1 + \epsilon)} )} $$

%         $$ 
%     \mathcal{R}^{ji} = -\underset{\mathbb{P} \in \mathcal{M}_{out}}{\inf}
% ~\mathbb{E}_\mathbb{P} \left[y_j | y_i \in \mathcal{W}_{\delta_i} \right] = h(\epsilon)  \hspace{5mm} \text{if} \hspace{5mm} \rho < 0 $$ 
%         $$ \mathcal{R}^{ji} = -\underset{\mathbb{P} \in \mathcal{M}_{out}}{\inf}
% ~\mathbb{E}_\mathbb{P} \left[y_j | y_i \in \mathcal{W}_{\delta_i} \right] = h(-\epsilon)  \hspace{5mm} \text{if} \hspace{5mm} \rho > 0 $$ 

\[
    \mathcal{R}^{ji}: = \begin{cases}
    \frac{d}{h(-\epsilon)} -1, &\text{if}  ~  \rho_{ji} < 0 \\
 \frac{d}{h(\epsilon)} -1, &\text{if} ~ \rho_{ji} > 0 \\
 0,  &\text{if} ~ \rho_{ji} = 0 \\
\end{cases}
\]
where $h(\cdot)$ is defined as in\eqref{eq:h_eps}.
\end{theorem}

\begin{figure}[t]
    \begin{subfigure}[t]{.48\linewidth}
        \centering
    \includegraphics[width=\linewidth]{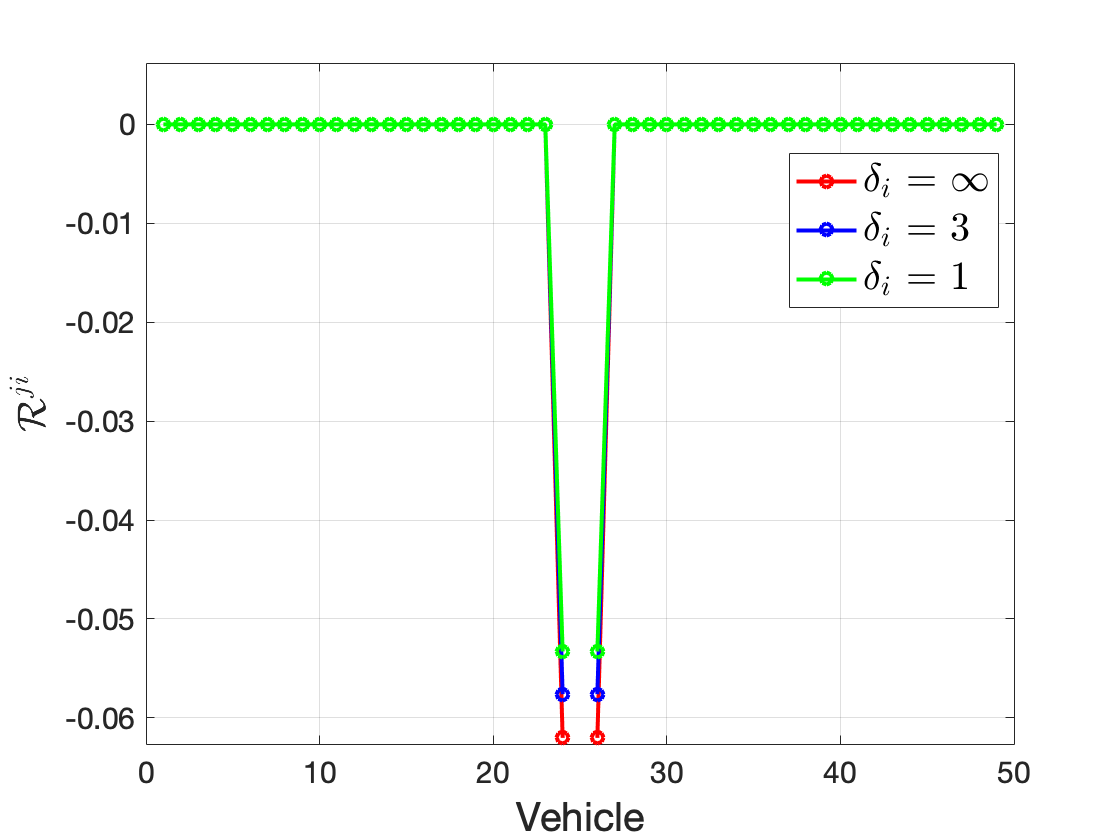}
    	\caption{The complete graph.}
    \end{subfigure}
    \hfill
    \begin{subfigure}[t]{.48\linewidth}
        \centering
    \includegraphics[width=\linewidth]{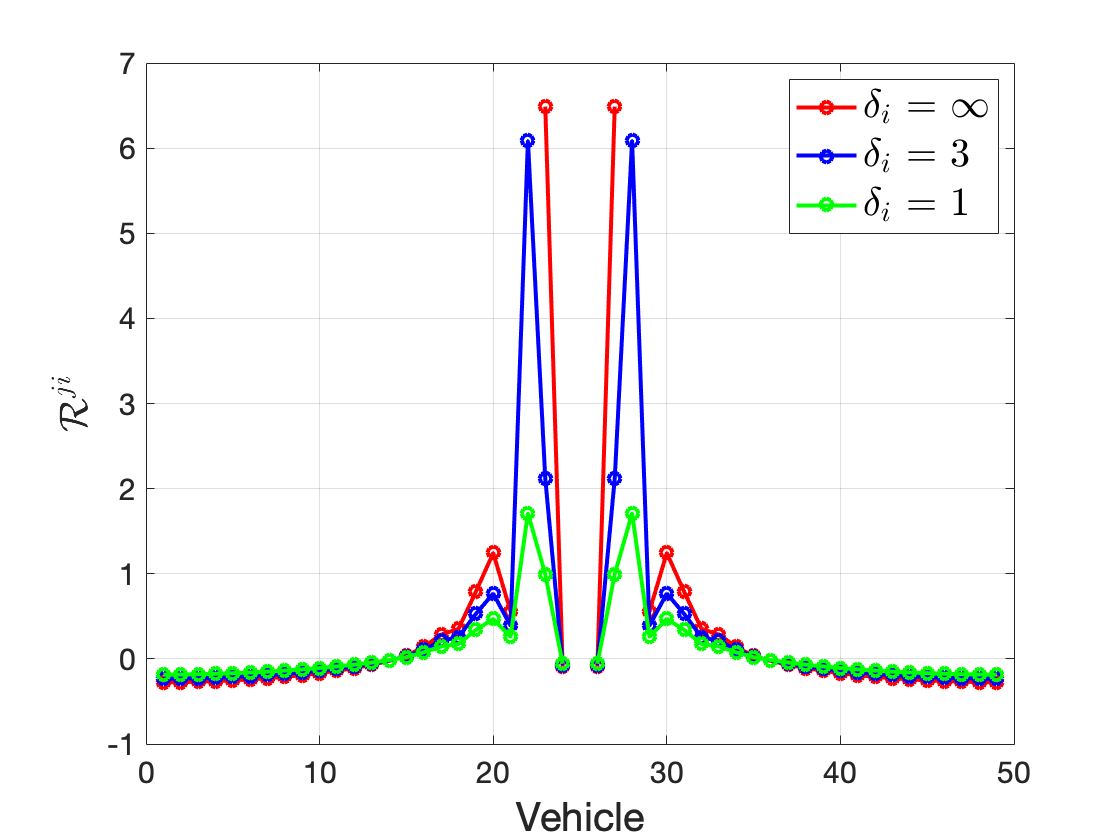}
    	\caption{The $6-$cycle graph.}
    \end{subfigure}
    \hfill
    % \begin{subfigure}[t]{.24\linewidth}
    %     \centering
    % 	\includegraphics[width=\linewidth]{Figs/mul_cas_ren_5-cycle_num.png}
    % 	\caption{The $5-$cycle graph.}
    % \end{subfigure}
    % \hfill
    \begin{subfigure}[t]{.48\linewidth}
        \centering
\includegraphics[width=\linewidth]{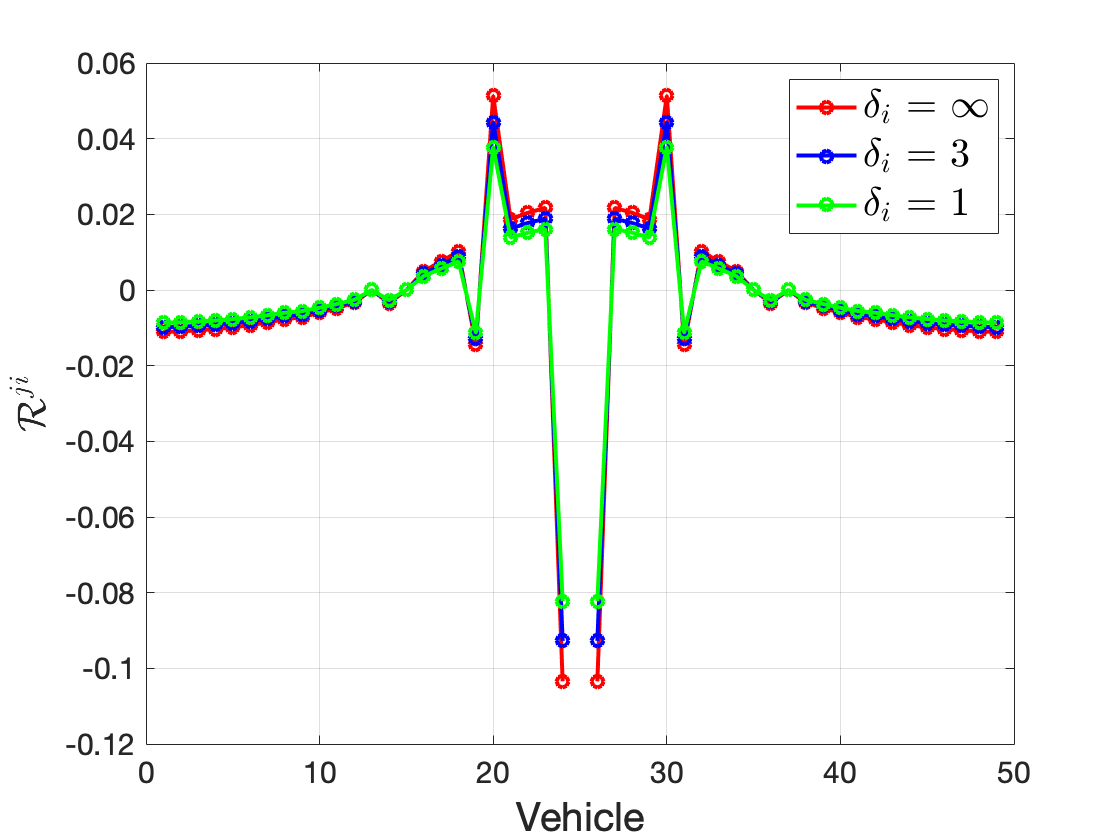}
    	\caption{The $10-$cycle graph.}
    \end{subfigure}
    \hfill
    \begin{subfigure}[t]{.48\linewidth}
        \centering
    \includegraphics[width=\linewidth]{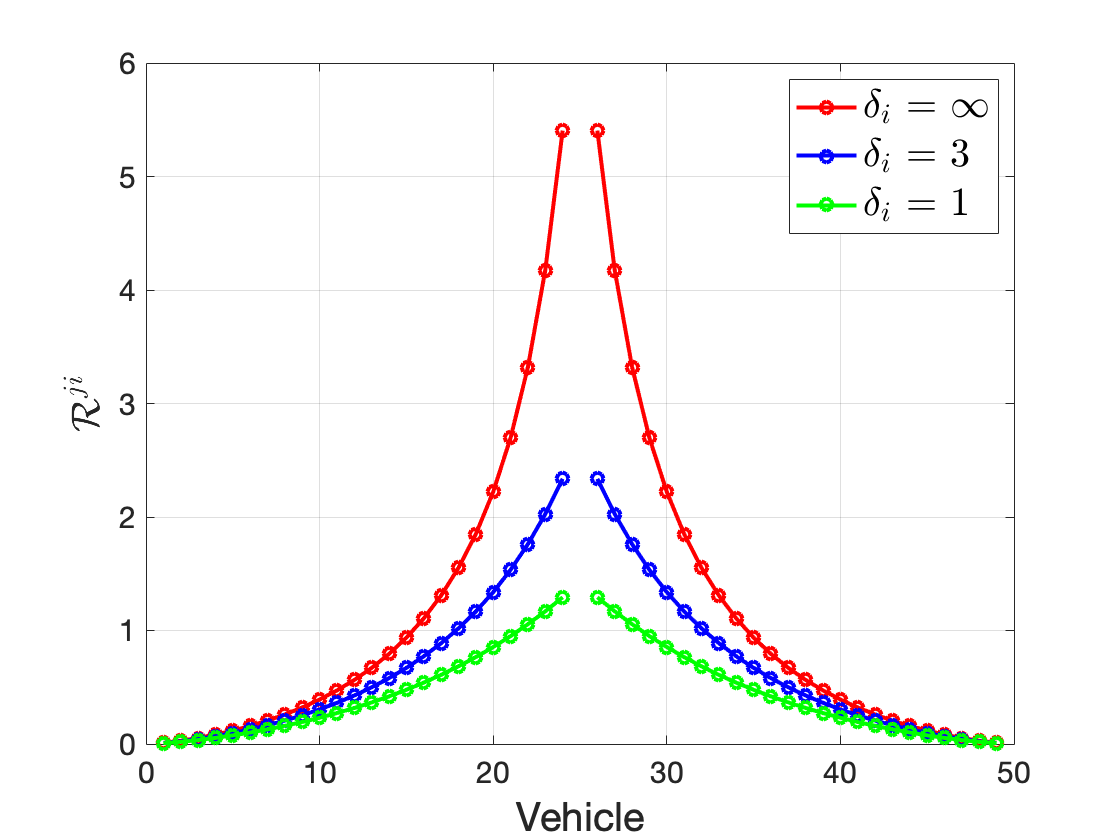}
    	\caption{The path graph.}
    \end{subfigure}
    \hfill
    \caption{The distributionally robust cascading risk profile for different value of $\delta_i$ for various graph topologies. }
    \label{fig:risk_di}
\end{figure}

% \guangyi{Please resize the figure into the same size and increase the size of the font if possible}

\subsection{Bound on Distributionally Robust Cascading Risk}
We provide a lower bound on the distributionally robust cascading risk for the system   \eqref{eq:dyn_platoon_gen_cov} as a function of eigenvalues of the covariance matrix of the steady state inter-vehicle distance. This quantification can provide insight into the maximum possible safety of the platoon.
% and by extension a bound on  distributionally robust cascading risk for the system 

% perturbed by noise for a general covariance matrix .

% Consider the system
% \begin{equation} \label{eq:dyn_platoon_gen_cov}
% \begin{aligned}
%     d\bm{\text{x}}_t &= \bm{\text{v}}_t dt \\
%     d\bm{\text{v}}_t &=  -L \bm{\text{v}}_{t-\tau} dt - \beta L (\bm{\text{x}}_{t-\tau} - \bm{\text{y}}) dt + E d\bm{{\xi}_t}, 
% \end{aligned}
% \end{equation}
% The system governed by \eqref{eq:dyn_platoon_gen_cov} may arise if we consider communication noise in the control law. Unlike \eqref{eq:dyn_platoon} which has input covariance matrix, $\Gamma = g^2I$,
% \eqref{eq:dyn_platoon_gen_cov} has covariance matrix $\Gamma = EE^T$.

\begin{lemma}   \label{lem:drf_bound}
The distributionally robust cascading risk for the system
\ref{eq:dyn_platoon_gen_cov} satisfies the following lower bound
$$
\mathcal{R}^{ji}\geq \dfrac{d}{f(\mu_{n-1}, \mu_1)} - 1,
$$
where $f(\mu_{n-1}, \mu_1) = \dfrac{\sqrt{\mu_{n-1}(1 + \epsilon)}}{\sqrt{\frac{1}{2}\left(1 + \textup{erf}\left(\frac{(d_i^* - d)}{2 \sqrt{\mu_1 (1 - \epsilon)}}\right)\right)}}$
and $\mu_{n-1}$ and $\mu_1$ are the largest  and the smallest eigenvalue of $\tilde{\Sigma}_0$ respectively and $d_i^* \in \R_+$.
\end{lemma}

% $$
% \underset{\mathbb{P} \in \mathcal{M}_{out}}{\sup}\mathbb{E}_\mathbb{P}[\Tilde{d_j} | \Tilde{d_i} \in \mathcal{W}_{\delta_i}] \leq f(\mu_n),
% $$
% where $f(\mu_n) = \dfrac{\sqrt{\mu_n(1 + \epsilon)}}{\sqrt{\frac{1}{2}\left(1 + \textup{erf}\big(\frac{y_i^*}{2 \sqrt{\mu_n (1 + \epsilon)}}\big)\right)}},$

\section{Case Studies}\label{sec:case_studies}
\begin{figure}[t]
    \begin{subfigure}[t]{.48\linewidth}
        \centering
    \includegraphics[width=\linewidth]{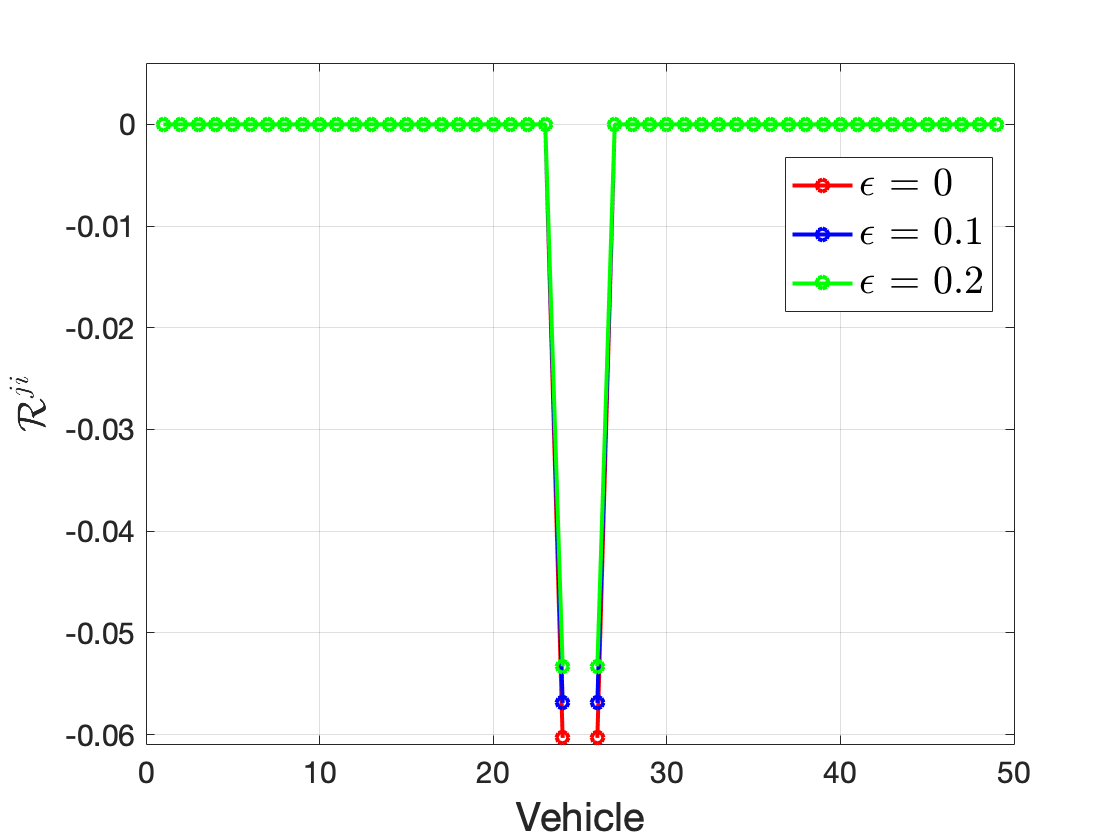 }
    	\caption{The complete graph.}
    \end{subfigure}
    \hfill
    \begin{subfigure}[t]{.48\linewidth}
        \centering
    \includegraphics[width=\linewidth]{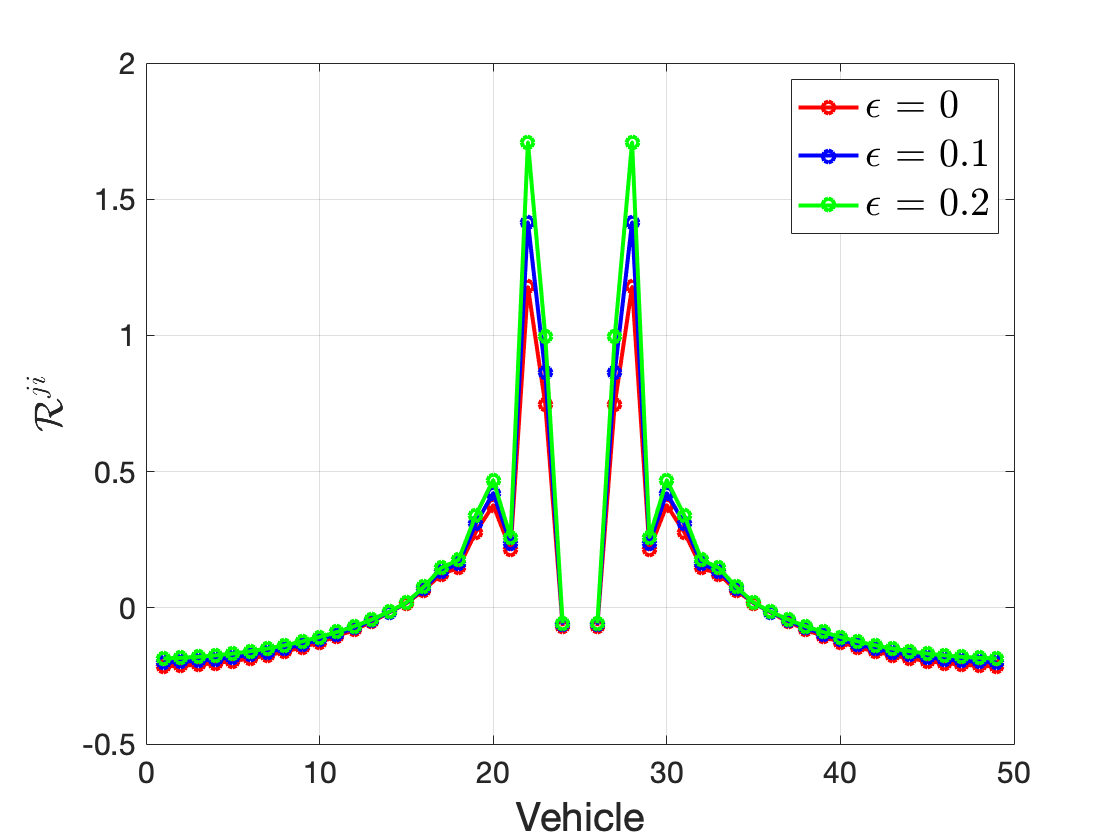}
    	\caption{The $6-$cycle graph.}
    \end{subfigure}
    \hfill
    % \begin{subfigure}[t]{.24\linewidth}
    %     \centering
    % 	\includegraphics[width=\linewidth]{Figs/mul_cas_ren_5-cycle_num.png}
    % 	\caption{The $5-$cycle graph.}
    % \end{subfigure}
    % \hfill
    \begin{subfigure}[t]{.48\linewidth}
        \centering
\includegraphics[width=\linewidth]{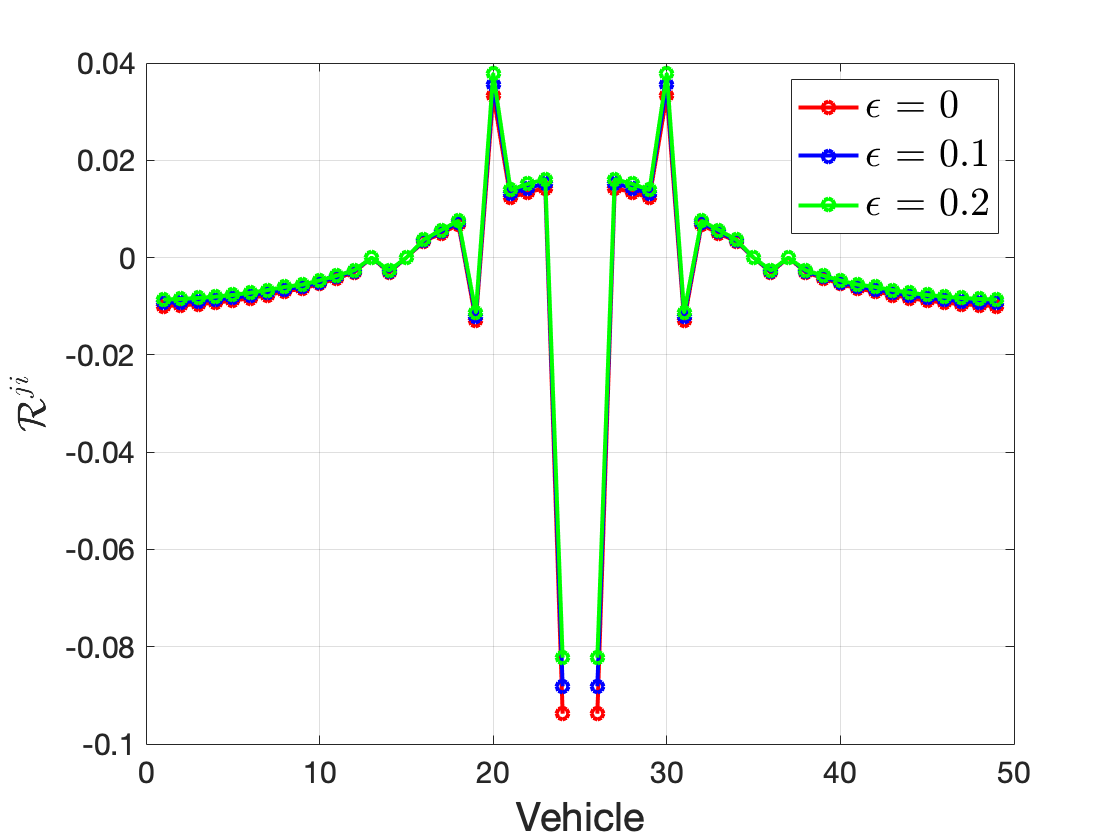}
    	\caption{The $10-$cycle graph.}
    \end{subfigure}
    \hfill
    \begin{subfigure}[t]{.48\linewidth}
        \centering
    \includegraphics[width=\linewidth]{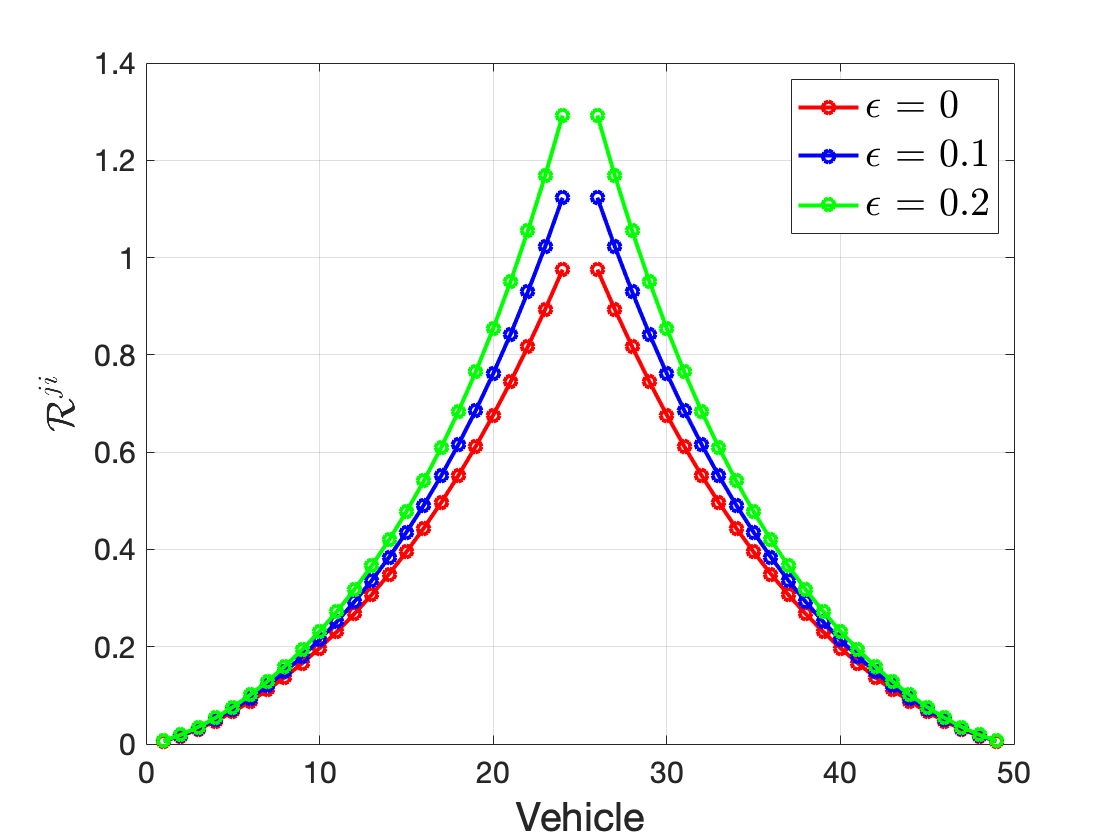}
    	\caption{The path graph.}
    \end{subfigure}
    \hfill
    \caption{The cascading risk profile conditioned on different value of $\epsilon$ for various graph topologies.}
    \label{fig:risk_epsilon}
\end{figure}

\begin{figure*}[t]
    \begin{subfigure}[t]{.24\linewidth}
        \centering
    \includegraphics[width=\linewidth]{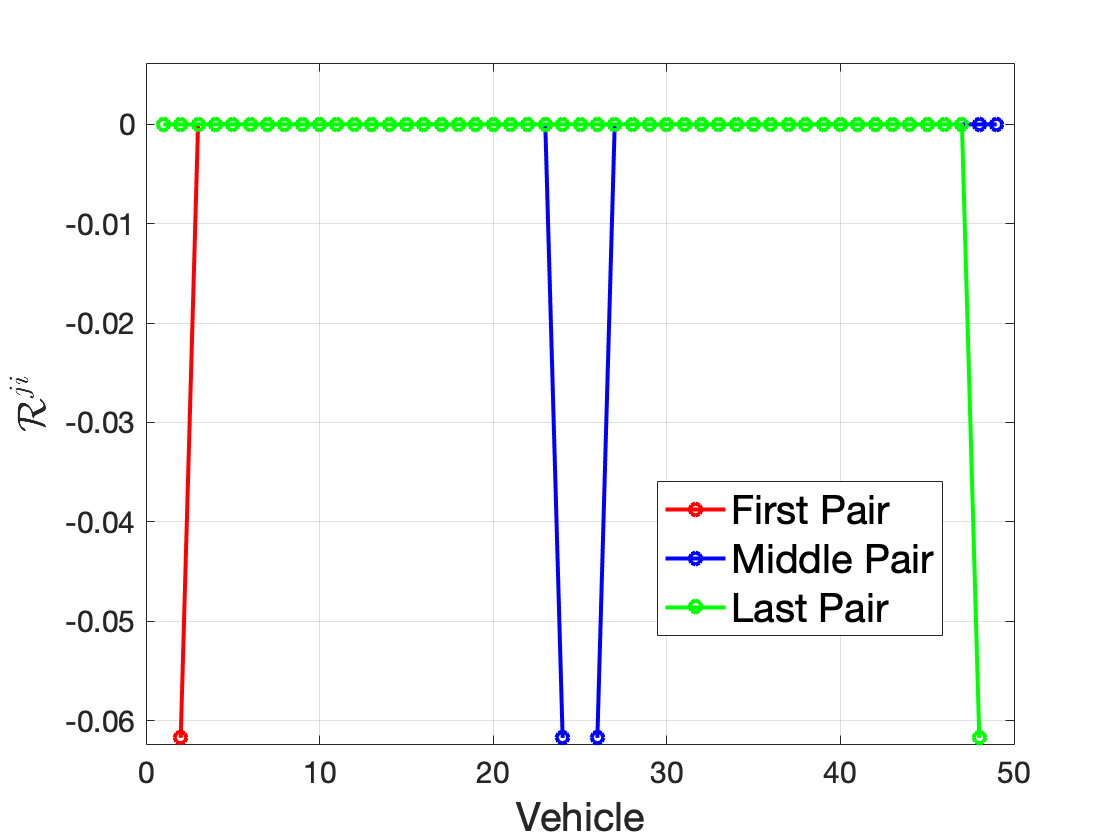}
    	\caption{The complete graph.}
    \end{subfigure}
    \hfill
    \begin{subfigure}[t]{.24\linewidth}
        \centering
    \includegraphics[width=\linewidth]{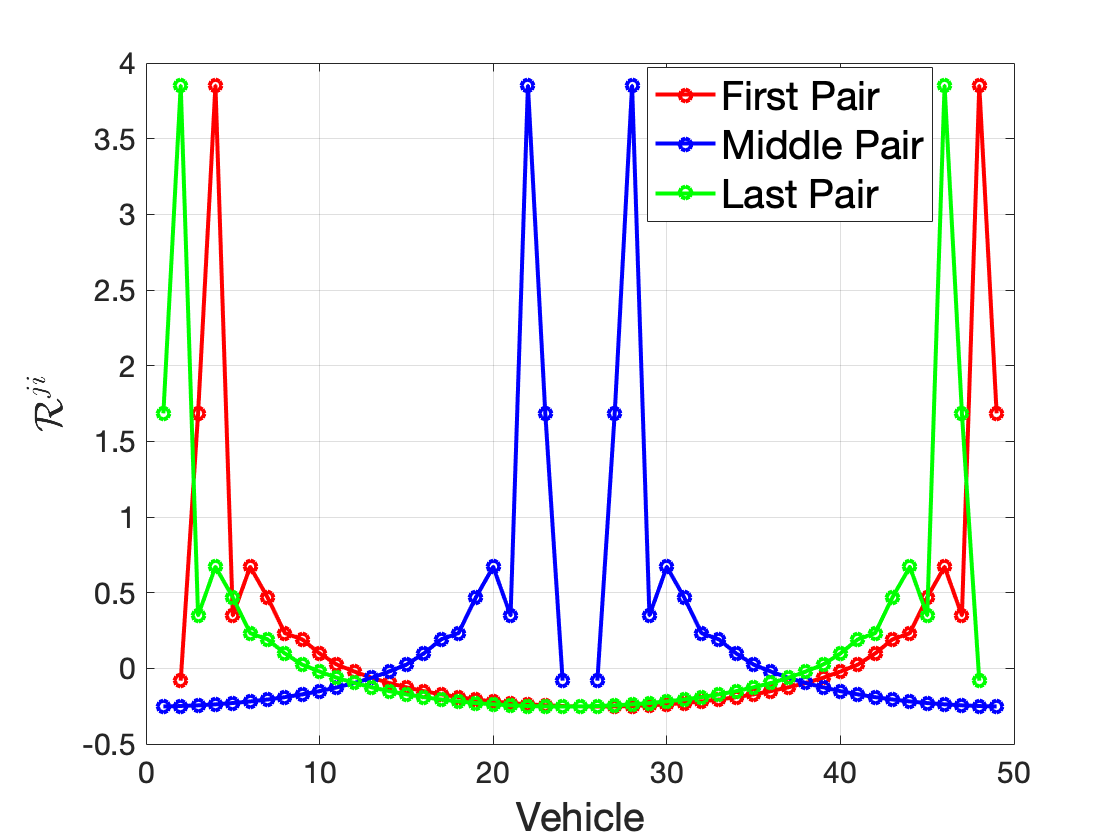}
    	\caption{The $6-$cycle graph.}
    \end{subfigure}
    \hfill
    % \begin{subfigure}[t]{.24\linewidth}
    %     \centering
    % 	\includegraphics[width=\linewidth]{Figs/mul_cas_ren_5-cycle_num.png}
    % 	\caption{The $5-$cycle graph.}
    % \end{subfigure}
    % \hfill
    \begin{subfigure}[t]{.24\linewidth}
        \centering
\includegraphics[width=\linewidth]{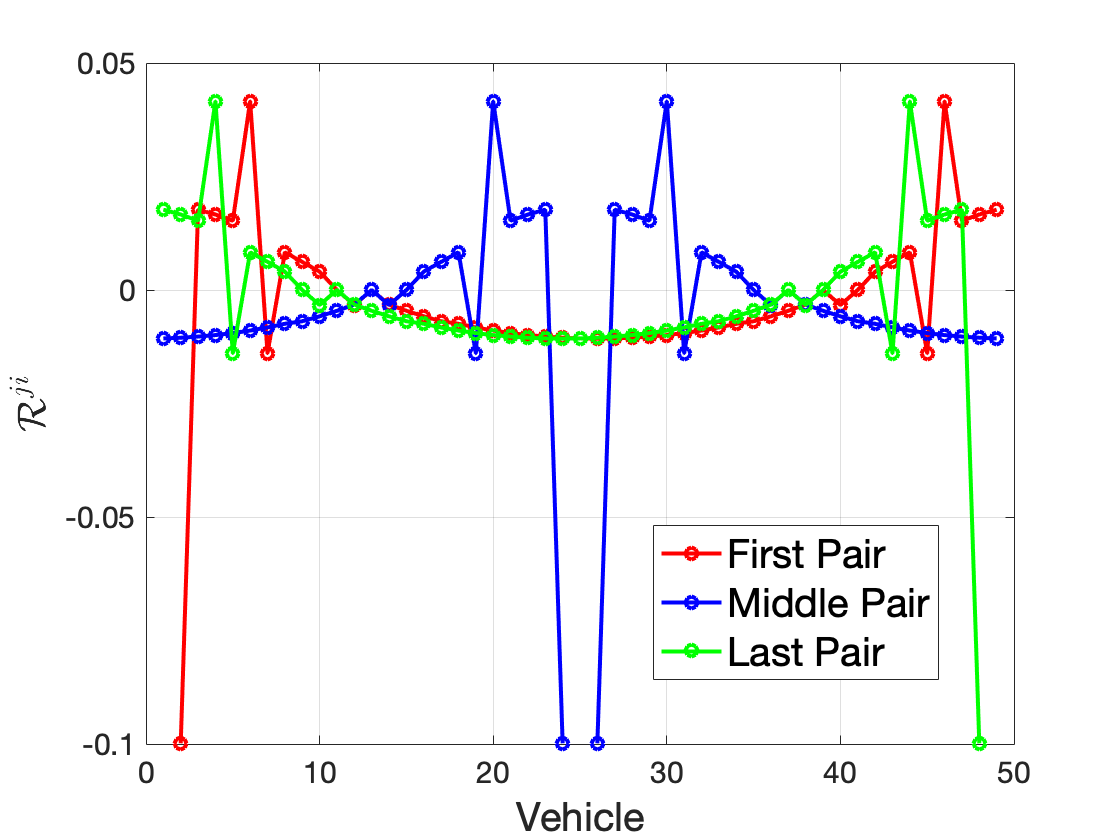}
    	\caption{The $10-$cycle graph.}
    \end{subfigure}
    \hfill
    \begin{subfigure}[t]{.24\linewidth}
        \centering
    \includegraphics[width=\linewidth]{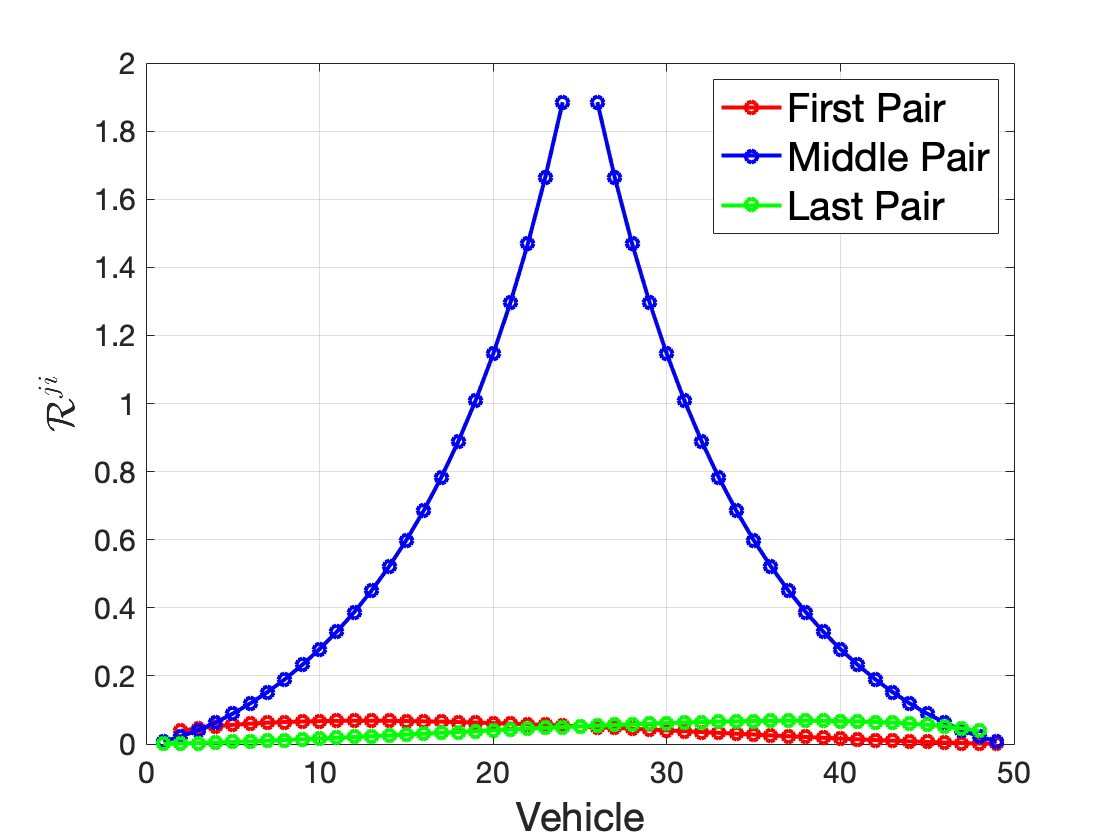}
    	\caption{The path graph.}
    \end{subfigure}
    \hfill

    \caption{The cascading risk profile at various location of the $i$'th pair for various graph topologies for $\delta_i = 3$.}.
    \label{fig:risk_location}
\end{figure*}

We discuss the case studies for the system of platoon \eqref{eq:dyn_platoon} for the communication network governed by the complete graph, the path graph and the p-cycle graphs. We present various cases depending on the location of the $i$'th pair in the platoon for different collection of supersets $\mathcal{W}_{\delta_i}$. We set the number of agents $n = 50$, the steady state inter-vehicle distance $d = 2$ and $c = 1$for all simulations. We set $\epsilon = 0.2$ for simulations in Fig. \ref{fig:risk_di}, which shows risk profile as a function of $\delta_i$ and Fig \ref{fig:risk_location}, where we show risk profile for different locations of the $i$'th pair in the platoon. The Fig \ref{fig:risk_epsilon} shows the risk profile for different values of $\epsilon$.\\
1) Complete Graph: We set $g_0 = 10, \beta = 1 $ and $\tau = 0.02$. As evident from Fig \ref{fig:risk_di}, the cascading effect is limited to the immediate neighbour of the location of the $i$'th pair. This is due to the result of the covariance matrix for the complete graph being a tri-diagonal matrix \cite{liu2021risk}. Hence, the correlation $\rho_{ji}$ for all pair $j$ such that $|j - i|>1$ equals $0$. This risk being zero for all pair $j$ for which $\rho_{ji}=0$ is also supported by \ref{thm: exptd_value}. For $\rho = 0$, the conditional expected value of a normal random variable equals the mean of the random variable, which is $d$ for the steady state inter-vehicle distance. We also see that the risk profile for the complete graph shifts depending upon the location of the $i$'th pair as shown in Fig. \ref{fig:risk_location} 

2) Path Graph: We set $g_0 = 0.25, \beta = 4 $ and $\tau = 0.05$. We see that the cascading risk profile decays as location of the $j$'th pair increases from the location of the $i$'th pair as shown in Fig. \ref{fig:risk_di} and Fig. \ref{fig:risk_location}. The magnitude of the risk  depends upon the location of $i$'th pair as shown in \ref{fig:risk_location}. The location of the $i$'th pair in the middle leads to a higher risk to neighbouring pair of vehicles compared to location of $i$'th pair at the start or the end of the platoon.  

2) $p$-Cycle Graph: We set $g_0 = 4, \beta = 2 $ and $\tau = 0.01$. We present the result for cascading risk profile for $p = 6$ and $p=10$. As expected, the risk profile converges towards the profile for the complete graph as $p$ increases. The location of the $i$'th pair plays a very important role here as we can see in Fig \ref{fig:risk_di} and Fig \ref{fig:risk_location}. When the $i$'th pair is in middle, we get a symmetric profile but when the $i$'th pair lies at either end, the cascading risk profile is more pronounced for vehicles at the start and end of the platoon while the vehicles in the middle of the platoon are relatively safe. 

The Fig. \ref{fig:risk_epsilon} outlines the importance of considering the ambiguity set of probability measures to quantify the risk since the risk is subject to steady state statistics which may be different for different probability measures in the ambiguity set. The monotonicity of the conditional expected value with $\epsilon$ as stated in \ref{thm: dr_risk} is evident in Fig \ref{fig:risk_epsilon}.

\section{Conclusion} \label{sec:conclusion}
This work is our first step towards developing a distributionally robust risk framework for the analysis of cascade of failures (e.g., collision or soft failures) in networked dynamical systems. We obtain explicit formulas to quantify ambiguity set and distributionally robust cascading risk. Both our theoretical results and simulations show how the change in steady state statistics of the input noise affects the risk profile of vehicles in the platoon emphasizing the importance of ambiguity set of probability measures in risk calculation. 

% inter-vehicle distance of one pair affects the safety of other pair in the platoon. 

Some interesting extensions of the current risk framework that we intend to explore in the future includes developing closed form distributionally robust cascading risk formulas by conditioning distributionally robust functional for multiple pair of inter-vehicle distance. Our future work also focuses on extending this framework for a wider range of ambiguity set of probability measures. Lastly, we would like to explore how network parameters such as connectivity and effective resistance impose a fundamental limit on minimum achievable risk for the networked system.  
\appendix
% \subsubsection{Proof of Lemma \ref{lem: amb_set_norm}}:
% The steady state covariance matrix $P$ can be written explicitly as 
% \begin{align*} 
%     P &= \int_{0}^{\infty} \Phi_i(s) I \Phi_i^T(s) ds
% \end{align*}

% Similarly, the steady state covariance matrix $\R$ can be written explicitly as 
% \begin{align*} 
% R &= \int_{0}^{\infty} \Phi_i(s) \Sigma \Phi_i^T(s) ds
% \end{align*}

% The output ambiguity set $M_{out} $ can be written in the following form 
% \begin{align*}
%     \|P - R\|_F 
%     &\leq  \int_{0}^{\infty}\|\Phi_i(s)(I   - \Sigma) \Phi_i^T(s)\|_F ds\\
%     &\leq  \int_{0}^{\infty}\|\Phi_i(s)\|_F^2 \|(I   - \Sigma)|_F ds\\
%     & =  \|(I   - \Sigma)|_F \int_{0}^{\infty}\|\Phi_i(s)\|_F^2 ds\\
%     & = \eta \int_{0}^{\infty}\|\Phi_i(s)\|_F^2 ds\\
% \end{align*}
% This leads to $\epsilon = \delta \int_{0}^{\infty}\|\Phi_i(s)\|_F^2 dt $.
% From the exponential stability of the system \eqref{eqn:dyn_sys_1_delay}, it may be possible to find a $\|\Phi_i(s)\|_F^2$ such that the integral is finite since \eqref{eqn:dyn_sys_1_delay} is integrable.
% \hfill$\square$

\subsubsection{Proof of Lemma \ref{lem:amb_set_cone}} 
Consider the system defined by \eqref{eq:dyn_platoon_gen_cov}. The solution of the decoupled version of this system can be written as 
$$
\begin{bmatrix}
    \tilde{\bm{z}}_t \\
    \tilde{\bm{v}}_t
\end{bmatrix} = \tilde{\Xi}(\tilde{\bm{z}}_{[-\tau,0]}, \tilde{\bm{v}}_{[-\tau,0]}, \tilde{\Phi}(t-s)) + \int_{0}^{t}\tilde{\Phi}\tilde{Q}E d\bm{{\xi}_s}, 
$$
where $\tilde{\Phi}$ is the principal solution of \eqref{eq:dyn_platoon_gen_cov}, both $\tilde{\Xi}(\cdot, \cdot, \cdot) $ and $\tilde{\Phi}$ are exponentially decaying and $\tilde{Q} = [\bm{0}_{n \times n}; Q^T] E$ \cite{somarakis2018risk}. Let $ \Tilde{\bm{\psi}}_t = [\tilde{\bm{z}}_t; \tilde{\bm{v}}_t ]$
The steady state covariance matrix for the system \eqref{eq:dyn_platoon_gen_cov} can be written as  
\begin{align*} 
    \mathbb{E}_{\mathbb{P}}[\bar{\Tilde{\bm{\psi}}} \, \bar{\Tilde{\bm{\psi}}}^T] &= \int_{0}^{\infty} \tilde{\Phi}(s)\Bar{Q}E E^T  \Bar{Q}^T \tilde{\Phi}^T(s)ds\\
    &=\int_{0}^{\infty} \tilde{\Phi}(s)\Bar{Q}\tilde{\Gamma}\Bar{Q}^T \tilde{\Phi}^T(s)ds\\
    & \preceq (1 + \epsilon)\int_{0}^{\infty} \tilde{\Phi}(s)\Bar{Q}\tilde{\Gamma}_0 \Bar{Q}^T \tilde{\Phi}^T(s)ds
\end{align*}
Using a linear transformation $ \tilde{\bm{z}}_t = T\Bar{\bm{\psi}}_t$ \cite{somarakis2018risk}, where $T$ is a $n \times 2n$ matrix with $p$'th row being $1 \times 2n$ canonical vector $\bm{e}^T_{2p -1}$, leads to  
$$\mathbb{E}_{\mathbb{P}}[\bar{\tilde{\bm{z}}} \bar{\tilde{\bm{z}}} ^T]  = T \mathbb{E}_{\mathbb{P}}[\bar{\Tilde{\bm{\psi}}} \bar{\Tilde{\bm{\psi}}}^T] T^T. $$
Now using the transformation similar to \eqref{eq:sigma_transform} leads to $\Tilde{\Sigma} \preceq (1 + \epsilon) \Tilde{\Sigma}_0$,
where
$$
\Tilde{\Sigma}_0 = D^T QT\bigg(\int_{0}^{\infty} \tilde{\Phi}(s)\Bar{Q}\Gamma_0 \Bar{Q}^T \tilde{\Phi}(s)^Tds\bigg) T^T Q^T D.
$$
The other inequality follows similarly.
\hfill$\square$

\subsubsection{Proof of Theorem \ref{thm: exptd_value}}
For simplicity, let $\rho = \rho_{ji}$ and $\rho' = \sqrt{1 - \rho_{ji}^2}.$ We write the conditional expected value as 
\begin{align*}
     \mathbb{E}_\mathbb{P}[\bar{d}_j| \bar{d}_i \in \mathcal{W}_{\delta_i}]=  \frac{\mathbb{E}_\mathbb{P}[\bar{d}_j \bm{1}_{[\bar{d}_i \in \mathcal{W}_{\delta_i}]}]}{\mathbb{E}_\mathbb{P}[\bm{1}_{[\bar{d}_i \in \mathcal{W}_{\delta_i}]}]}
\end{align*}
Using the bi-variate normal distribution, we write the RHS of the above equation as  
\begin{align*}
        \dfrac{\int_{-\infty}^{\infty}\int_{-\infty}^{d_i^*}\frac{\bar{d}_je^{-\frac{1}{2 \rho'^2}\big[ \rho'^2 \frac{(\bar{d}_j - d)^2}{\sigma_j^2} + \big(\frac{(\bar{d}_i - d)}{\sigma_i} - \rho \frac{(\bar{d}_j -d)}{\sigma_j}\big)^2\big]} d\bar{d}_i d\bar{d}_j}{2\pi \sigma_j\sigma_i \rho'} }{\int_{-\infty}^{d_i^*} \frac{1}{\sqrt{2 \pi} \sigma_i} e^{-\big(\frac{(\bar{d}_i -d)}{\sigma_i}\big)^2}}.
\end{align*}

We use Fubini's theorem \cite{durrett2019probability} to split the double integral into iterated integral. It is trivial to verify that the conditions for Fubini's Theorem are satisfied by the integrand. Integrating with respect to $\bar{d}_i$ yields
    \begin{align*}
    \dfrac{\int_{-\infty}^{\infty}\bar{d}_j \frac{e^{-\frac{1}{2 } \frac{(\bar{d}_j - d)^2}{\sigma_j^2}}}{\sqrt{2 \pi} \sigma_j}  \frac{1}{2}\big[1 + \textup{erf}\big[ \frac{d_i^* - \rho \frac{\sigma_i}{\sigma_j}\bar{d}_j}{\sqrt{2}\rho'\sigma_i}\big] \big] d\bar{d}_j}{\frac{1}{2}\big[1 + \textup{erf}\big[ \frac{(d_i^*-d)}{\sqrt{2}\sigma_i}\big] \big]}.
\end{align*}
It is straightforward to see that the first term in numerator simplifies to
 $$\frac{1}{2}\int_{-\infty}^{\infty}\bar{d}_j \frac{e^{-\frac{(\bar{d}_j - d)}{2\sigma_j^2}^2}}{\sqrt{2 \pi} \sigma_j} d\bar{d}_j = \frac{d}{2},$$
 % in the numerator equals $d$ 
 as the integral results in the expected value of normal random variable with mean $d$. 
 % In order to write the algebraic expressions in compact form, we introduce following notations, $\alpha_1 = \frac{1}{2\sigma_j^2}$, $\beta = \frac{y^*}{\sqrt{2}\rho' \sigma_i}$, $\alpha_2 = \frac{\rho}{\sqrt{2}\rho' \sigma_j}$
%  \begin{align*}
%      \mathbb{E}_Q[y_j| y_i \in \mathcal{W}_i]
%     &= \frac{\frac{\int_{-\infty}^{\infty}y_j e^{- a_1 y_j^2} erf(\beta - a y_j)}{\sqrt{2 \pi} \sigma_j}dy_j}{\big[1 + erf\big[ \frac{y^*}{\sqrt{2}\sigma_i}\big] \big]}
% \end{align*}
% Using the following result for $a_1 > 0, a\geq 0$ \cite{korotkov2020_error}
The second term in the numerator is
$$
\int_{-\infty}^{\infty}\bar{d}_j \frac{e^{-\frac{1}{2 } \frac{(\bar{d}_j - d)^2}{\sigma_j^2}}}{\sqrt{2 \pi} \sigma_j}  \frac{1}{2}\textup{erf}\bigg[ \frac{d_i^* -d(1 - \rho\frac{\sigma_i}{\sigma_j}) - \rho \frac{\sigma_i}{\sigma_j}\bar{d}_j}{\sqrt{2}\rho'\sigma_i}\bigg] d\bar{d}_j,
$$ which, after some algebraic manipulations simplifies to  
$$
e^{\frac{- d^2}{2\sigma_j}}
\int_{-\infty}^{\infty}\frac{\bar{d}_j e^{ \frac{(-\bar{d}_j^2 + 2 d\bar{d}_j)}{2\sigma_j^2}} \textup{erf}\bigg[ \frac{d_i^* -d(1 - \rho\frac{\sigma_i}{\sigma_j}) - \rho \frac{\sigma_i}{\sigma_j}\bar{d}_j}{\sqrt{2}\rho'\sigma_i}\bigg]}{{2\sqrt{2 \pi}\sigma_j}} d\bar{d}_j.
$$
We solve this integral using the following integral solution in \cite{korotkov2020_error} 

\begin{multline*}
\int_{-\infty}^{\infty}z e^{(-a_1 z^2 + b_1 z)} \textup{erf}(a_2z + b)dz = \\\frac{\sqrt{\pi}b_1}{2\pi a\sqrt{a}}e^{\frac{b_1^2}{4a_1}} A + \frac{a_2}{a_1\sqrt{a_1 + a_2^2}}B,
\end{multline*}
where $A = \textup{erf}\bigg(\frac{2a_1b_2 +a_2b_1}{2 \sqrt{a_1^2 + a_1 a_2^2}}\bigg)$ and $B = e^{\frac{b_1^2 - 4a_1b_2^2 - 4a_2b_1b_2}{4a_1 +4a_2^2}}. $

We substitute $a_1=\frac{1}{2\sigma_j^2}, b_1=\frac{d}{2\sigma_j^2}$ and $ a_2=\frac{-\rho}{\sqrt{2}\sigma_j \rho'}, b_2 = \frac{d_i^* -d(1 - \rho\frac{\sigma_i}{\sigma_j}) }{\sqrt{2}\sigma_j \rho'}$. 
The result follows after some algebraic manipulation and putting all the pieces of solution together.

% \begin{multline}     \label{eq:control_law}
%        {u}_t^{(i)} = \sum_{j = 1}^{n}k_{ij}\big(v^{j}_{t - \tau} - v^{i}_{t - \tau}\big)
%     \\+ \beta \sum_{j = 1}^{n}k_{ij}\big(x^{j}_{t - \tau} - x^{i}_{t - \tau} -(d_j - d_i)\big).
% \end{multline}
% $$
% \int_{-\infty}^{\infty}z e^{- a_1 z^2} erf(a z + \beta)dz =\frac{a}{a_1 \sqrt{a^2 + a_1}}e^{-\frac{a_1 \beta^2}{a^2 + a_1}} 
% $$
% It is straightforward to see that in our case , coefficient $-a>0$ for $\rho <0$. The result follows by simple algebraic calculation. For $\rho\geq0$, we use the property of error function being odd to get the coefficient $a \geq 0$. The result then follows by similar calculations.
 
 \subsubsection{Proof of Theorem \ref{thm: dr_risk}}

 Firstly, we prove the monotonicity of the functional 
$$\mathbb{E}_{\mathbb{P}}[\bar{d}_j | \bar{d}_i \in \mathcal{W}_{\delta_i}] = d -\sqrt{\frac{{2}}{{\pi}}}\frac{\rho_{ji} \sigma_j e^{\left(-\frac{(d_i^* -d)^2}{2 \sigma_i^2}\right)}}{1 + \textup{erf}\left(\frac{d_i^* - d}{\sqrt{2} \sigma_i}\right)},
$$
with respect to $g$. From \eqref{eq:steady_state_cov}, we know that $\sigma_{ij}$ is a function of $g$. We write $\sigma_i = \beta g$ and $\sigma_j = \alpha g$, where $\alpha, \beta > 0$.  For notational simplicity, we also let $A_1(g) = \bigg( {1 + \textup{erf}\left(\frac{d_i^* - d}{\sqrt{2} \sigma_i}\right)} \bigg) $ and $A_2(g) = e^{-\frac{(d_i^* -d)^2}{2 \beta^2 g^2}}$. Differentiating the functional with respect to $g$ leads to 

   % \frac{d}{dg}(\mathbb{E}_{\mathbb{P}}[\bar{d}_j | \bar{d}_i \in \mathcal{W}_{\delta_i}]) = 
$$
-\alpha \rho_{ji}A_2(g)\bigg(A_1(g) \bigg(1+\frac{(d_i^* -d)^2}{\beta^2 g^2}\bigg)+ A_2(g) \frac{(d_i^* -d)^2}{\beta^2 g^2}\bigg) 
$$
It is straighforward to verify that $A_1(g) > 0$ since it is twice the probability of the random variable $\bar{d}_i$ with mean $d$ and variance $\sigma_i.$ Similarly, $A_2(g) > 0$ since it is an exponential function.  \\
All the terms in the derivative is positive except for $\rho_{ji}$, which can be positive, negative or zero.
For $\rho_{ji} = 0$, $\mathbb{E}_{\mathbb{P}}[\bar{d}_j | \bar{d}_i \in \mathcal{W}_{\delta_i}] = d$ for all values of $g$. For $\rho_{ji} > 0$, the derivative is negative which results in $\mathbb{E}_{\mathbb{P}}[\bar{d}_j | \bar{d}_i \in \mathcal{W}_{\delta_i}]$ being monotonically decreasing with $g$. Similarly, for $\rho_{ji} < 0$, $\mathbb{E}_{\mathbb{P}}[\bar{d}_j | \bar{d}_i \in \mathcal{W}_{\delta_i}]$ being monotonically increasing with $g$. As a result, we have 

\[
    \underset{\mathbb{P} \in \mathcal{M}_g}{\inf}\mathbb{E}_{\mathbb{P}}[\bar{d}_j | \bar{d}_i \in \mathcal{W}_{\delta_i}] = \begin{cases}
    {h(-\epsilon)}, &\text{if}  ~  \rho < 0 \\
 {h(\epsilon)}, &\text{if} ~ \rho > 0 \\
 d,  &\text{if} ~ \rho = 0 \\
\end{cases}
\]

The result follows by substituting $ \underset{\mathbb{P} \in \mathcal{M}_g}{\inf}\mathbb{E}_{\mathbb{P}}[\bar{d}_j | \bar{d}_i \in \mathcal{W}_{\delta_i}]$ in \eqref{eq:risk_defn_platoon}.

\subsubsection{Proof of Lemma \ref{lem:drf_bound}} In order to derive the bound on the distributionally robust risk, we first find a bound on conditional expected value as shown below
$$ 
     \mathbb{E}_\mathbb{P}[\bar{d}_j| \bar{d}_i \in \mathcal{W}_{\delta_i}]=  \frac{\mathbb{E}_\mathbb{P}[\bar{d}_j \bm{1}_{[\bar{d}_i \in \mathcal{W}_{\delta_i}]}]}{\mathbb{E}_\mathbb{P}[\bm{1}_{[\bar{d}_i \in \mathcal{W}_{\delta_i}]}]}.
$$
Using the Cauchy - Schwartz Inequality \cite{durrett2019probability}, 
\begin{align*}
     \mathbb{E}_\mathbb{P}[\bar{d}_j| \bar{d}_i \in \mathcal{W}_{\delta_i}]
     &\leq  \frac{\mathbb{E}_\mathbb{P}[\,\bar{d}_j^2\, ]^{\frac{1}{2}} \, \mathbb{E}_\mathbb{P} [ \bm{1}_{[\bar{d}_i \in \mathcal{W}_{\delta_i}]} ]^{\frac{1}{2}}}{\mathbb{E}_\mathbb{P}[\bm{1}_{[\bar{d}_i \in \mathcal{W}_{\delta_i}]}]}\\
     &=  \frac{\tilde{\sigma}_j}{\mathbb{P}[{\bar{d}_i \in \mathcal{W}_{\delta_i}}]^{\frac{1}{2}}}.
\end{align*}
Since $\tilde{\Sigma} \in S^{n-1}_{++}$ being a covariance matrix is symmetric, we use Schur-Horn Lemma \cite{horn1954stochastic, marshall2009_inequlities} to bound the diagonal elements of $\tilde{\Sigma}$ by its smallest and the largest eigenvalues which we denote by $\nu_1$ and $\nu_{n-1}$ respectively. This results in 
$$
\nu_1 \leq \tilde{\sigma}_k^2 \leq \nu_{n-1}
$$
for all $k \in \{1, \cdots, n\}$.
Using the above bound for $\sigma_j$ and $\sigma_i$
  \begin{align*}
     \mathbb{E}_\mathbb{P}[\bar{d}_j| \bar{d}_i \in \mathcal{W}_{\delta_i}]
     &\leq\frac{\sqrt{\nu_{n-1}}} {\sqrt{\frac{1}{2}\big(1 + \textup{erf}\big(\frac{(d^*- d)}{2 \sqrt{\nu_1}}\big)\big)}}.
 \end{align*}
It is straightforward to visualize that the expression on the RHS increases as $\nu_{n-1}$ increases and $\nu_1$ decreases. Consider $p_{n-1}$ to be the eigenvector corresponding to $\nu_{n-1}$. This results in
 % \begin{align*}
 %     p_{n-1}^T\tilde{\Sigma} p_{n-1} &\leq  (1 + \epsilon) p_{n-1}^T\tilde{\Sigma}_0 p_{n-1}^T\\
 %     \nu_{n-1} &\leq (1 + \epsilon) \underset{q_n \in \mathbb{R}^n}{\sup}p_{n-1}^T\tilde{\Sigma}_0 p_{n-1}^T\\
 %        &= \mu_{n-1}(1+\epsilon),
 % \end{align*}

 \begin{align*}
     p_{n-1}^T\tilde{\Sigma} p_{n-1} &\leq  (1 + \epsilon) p_{n-1}^T\tilde{\Sigma}_0 p_{n-1}^T\\
     \nu_{n-1} &\leq (1 + \epsilon) \underset{p \in \mathbb{R}^{n-1}}{\sup}p^T\tilde{\Sigma}_0 p^T\\
        &= \mu_{n-1}(1+\epsilon),
 \end{align*}

 where the last equality follows from the fact that the supremum of Rayleigh quotient results in the largest eigenvalue \cite{strang2017linearalgebra}. The same analysis for $\nu_1$ results in
  \begin{align*}
     \nu_1 
        \geq \mu_1(1 -\epsilon).
 \end{align*}

  Replacing $\nu_n$ and $\nu_1$ by $\mu_n(1+\epsilon)$ and $\mu_1(1 -\epsilon)$ in the expression of expected value leads to 
\begin{align*}
  {\underset{\mathbb{P} \in \mathcal{M}_{out}}{{\sup}} ~ \mathbb{E}_{\mathbb{P}}}[\bar{d}_j| \bar{d}_i \in \mathcal{W}_{\delta_i}] &= 
\dfrac{\sqrt{\mu_{n-1}(1 + \epsilon)}}{\sqrt{\frac{1}{2}\left(1 + \textup{erf}\left(\frac{(d_i^* - d)}{2 \sqrt{\mu_1 (1 - \epsilon)}}\right)\right)}}\\
&=  f(\mu_{n-1}, \mu_1)
\end{align*}
Finally, we find a lower bound as follows
\begin{align*}
    \mathcal{R}^{ji} &= \frac{d}  {\underset{{\mathbb{P}} \in \mathcal{M}_{out}}{{\inf}} ~ \mathbb{E}[\bar{d}_j | \bar{d}_i \in \mathcal{W}_{\delta_i}]} -1\\
    &\geq \frac{d}  {\underset{{\mathbb{P}} \in \mathcal{M}_{out}}{{\sup}} ~ \mathbb{E}[\bar{d}_j | \bar{d}_i \in \mathcal{W}_{\delta_i}]} -1
\end{align*}
The result follows by substituting the value of $  {\underset{\mathbb{P} \in \mathcal{M}_{out}}{{\sup}} ~ \mathbb{E}_{\mathbb{P}}}[\bar{d}_j| \bar{d}_i \in \mathcal{W}_{\delta_i}].$
 \hfill$\square$
     %       \underset{\mathbb{P} \in \mathcal{M}_{out}}{\sup}\mathbb{E}_\mathbb{P}[y_j| y_i \in \mathcal{W}_{\delta_i}]
     % &\leq\underset{\mathbb{P} \in \mathcal{M}_{out}}{\sup}\frac{\mu_n} {\sqrt{\frac{1}{2}\big(1 + erf\big(\frac{y_j}{2 \mu_n}\big)\big)}}

% Since, $\sigma_i$ and $\sigma_j$ varies with $g^2$, it is straightforward to see  $$\mathbb{E}_\mathbb{Q}[y_j | y_i \in \mathcal{W}_{\delta_i}] = \frac{|\rho|}{\sqrt{2 \pi}}\frac{\sigma_j e^{\frac{-\rho^2y^{*2}}{2 \sigma_i^2}}}{1 + erf(\frac{y^*}{\sqrt{2} \sigma_i})} $$ 

% is monotonic function in $g$ depending on sign of  $\rho$. 
% The result then follows by substituting the smallest or the largest value of $g = g_0 (1 + \epsilon)$ depending on $\rho$. The bound follows similarly.
%%%%%%%%%%%%%%%%%%%%%%%%%%%%%%%%%%%%%%%%%%%%%%%%%%%%%%%%%%%%%%%%%%%%%%%%%%%%%%%%%%%%%%%%%%%%%%
%END OF THE MAIN DOCUMENT
%%%%%%%%%%%%%%%%%%%%%%%%%%%%%%%%%%%%%%%%%%%%%%%%%%%%%%%%%%%%%%%%%%%%%%%%%%%%%%%%%%%%%%%%%%%%%%

\printbibliography
\end{document}